\documentclass[showkeys,reprint,twocolumn,floatfix,nofootinbib]{revtex4-1}

\usepackage[utf8]{inputenc}

\usepackage{amsmath,amssymb}

\usepackage{xcolor}
\definecolor{link-color}{RGB}{55,126,184}
\usepackage[colorlinks=true,%
            linkcolor   = {link-color},%
            citecolor   = {link-color},%
            urlcolor    = {link-color}]{hyperref}

\usepackage[]{graphicx}

\usepackage{booktabs}
\AtBeginDocument{
\heavyrulewidth=.08em
\lightrulewidth=.05em
\cmidrulewidth=.03em
\belowrulesep=.65ex
\belowbottomsep=0pt
\aboverulesep=.4ex
\abovetopsep=0pt
\cmidrulesep=\doublerulesep
\cmidrulekern=.5em
\defaultaddspace=.5em
}
\usepackage{tabularx}

\usepackage{siunitx}

\newcommand{\usd}[1]{\SI{#1}[\$\ensuremath{\,}]{}}
\newcommand{\AIC}{\ensuremath \text{AIC}}

\graphicspath{{figures/}{SI/figures/}}

\begin{document}

\title{Complex Contagion of Campaign Donations}
\date{\today}
\author{V.A. Traag}
\email{v.a.traag@cwts.leidenuniv.nl}
\affiliation{Royal Netherlands Institute of Southeast Asian and Caribbean Studies, Leiden, Netherlands}
\affiliation{e-Humanities Group, Royal Netherlands Academy of Arts and Sciences, Amsterdam, Netherlands}
\affiliation{CWTS, Leiden University, Leiden, Netherlands}

\begin{abstract}
Money is central in US politics, and most campaign contributions stem from a tiny, wealthy elite.
Like other political acts, campaign donations are known to be socially contagious.
We study how campaign donations diffuse through a network of more than $50\,000$ elites and examine how connectivity among previous donors reinforces contagion.
We find that the diffusion of donations is driven by independent reinforcement contagion: people are more likely to donate when exposed to donors from different social groups than when they are exposed to equally many donors from the same group.
Counter-intuitively, being exposed to one side may increase donations to the other side.
Although the effect is weak, simultaneous cross-cutting exposure makes donation somewhat less likely.
Finally, the independence of donors in the beginning of a campaign predicts the amount of money that is raised throughout a campaign.
We theorize that people infer population-wide estimates from their local observations, with elites assessing the viability of candidates, possibly opposing candidates in response to local support.
Our findings suggest that theories of complex contagions need refinement and that political campaigns should target multiple communities.
\end{abstract}

\keywords{complex contagion; social influence; campaign donation}

\maketitle

\section{Introduction}

No money, no campaign is a truism in US politics.
Financing a campaign is vital for any candidate to get elected.
Contrary to appearances, most donations come from individuals rather than corporations. 
While Obama in 2008 attracted the most donors ever, reaching over $300\,000$ people~\cite{lipsitz_filled_2011}, this still pales in comparison to the US population of over $300$ million.
The elite thus have a disproportionately large influence on campaigns and politics generally~\cite{gilens_testing_2014}.

The decision to donate or not, is not a purely individual one and is embedded in a wider social context~\cite{Verba1995}.
People who are asked to contribute are much more likely to do so~\cite{schlozman_civic_1999}.
Unsurprisingly, the wealthy are the most likely to donate~\cite{schlozman_civic_1999}.
Most money comes from large donations (i.e. $\geq \usd{200}$), which must be registered with the Federal Election Committee (FEC).
The FEC is a federal US agency tasked with the responsibility to administer and enforce campaign finance legislation, which is a contested subject in itself~\cite{kramer_randall_2007}.
Exacerbating the inequality in donations, the more wealthy and highly educated are also more likely to be asked to donate~\cite{brady_prospecting_1999,grant_give_2002} to the extent that when controlled for this selection bias, the effect of asking may even disappear~\cite{lim_mobilizing_2010}.
Such selection bias leads to an even greater inequality in financial contributions~\cite{schlozman_civic_1999}.
The wealthy and highly educated are also more likely to try to persuade others ~\cite{panagopoulos_contributions_2006}, while the influence is stronger from people who are emotionally close~\cite{brady_prospecting_1999}.
Research shows that individuals sitting on the same board donate to similar candidates, and that this effect also extends to members of interlocking boards~\cite{burris_interlocking_2005}.
Some scholars argued that spatial correlations in the geographical distribution of donations suggest they spread through social networks~\cite{gimpel_political_2006,tam_cho_contagion_2003}.
Personal contact is usually most effective for recruitment~\cite{Gerber2000}, while field experiments with canvassing also show contagious effects for turnout~\cite{nickerson_is_2008}. 

It is well known that most social contagion processes are not simple epidemic contagions in which a single contact is potentially sufficient for successful transmission of some virus~\cite{Centola2007,Karsai2014,Bakshy2012,Hodas2014,Weiss2014}.
For social contagion to materialize, social reinforcement is usually necessary, requiring multiple exposures to the behaviour before successful transmission~\cite{Granovetter1978}.
Many studies have analysed how the network structure affects diffusion of behaviour~\cite{Watts2002,Kearns2009,Centola2010,Bond2012,Bakshy2012,Banerjee2013,Coviello2014}, and different positions in the network may play different roles~\cite{Gonzalez-Bailon2011} during the adoption life cycle.
Nevertheless, little attention has been paid to how connectivity among adopters (i.e. donors) may affect contagion probabilities~\cite{Ugander2012}.
We here focus on the question how likely somebody is to adopt a certain behaviour (i.e. donating) given the adopters among his or her network neighbours, and pay particular attention to the connectivity among these adopting neighbours.
If somebody is exposed to adopters who are all acquainted with one another, the probability of that person adopting may be different from when none of the adopters know each other.
We here introduce two different modes in which structure may affect contagion: cohesive reinforcement and independent reinforcement.

If contagion is more likely when adopting neighbours know each other we will speak of cohesive reinforcement.
In this case, ego (the focal node) is embedded in a well-knit cohesive environment of adopters.
Such an environment is supportive and enables people to take risky actions, knowing they are supported by their friends~\cite{McAdam1986}.
It is also frequently a normative environment and part of the social influence is simply adhering to group norms~\cite{Sherif1936,Friedkin2001,Festinger1954}.
Failing to heed to the group's norms may lead to being ostracised, pressuring people to adopt the behaviour and follow the group norms~\cite{Elias1965}.
In the case of cohesive reinforcement contagion would be especially likely if adopters come from the same group of people.
Take for example protesting: when four friends participate but do not know each other, we may be less inclined to join than when these four friends know each other and we can participate as a group~\cite{Schussman2005}.
Such cohesive reinforcement may also play a role in collective action more generally.

If contagion is more likely when adopting neighbours don't know each other, we will speak of independent reinforcement.
In this setting, adopters from the same group do not reinforce social contagion.
In some way, signals from the same group are redundant, which is consistent with the strength of weak ties~\cite{Granovetter1973}: news passes through weak links, crossing group boundaries.
Information from multiple sources is generally more credible~\cite{Harkins1981}, but only if they are independent~\cite{Wilder1977,Harkins1987}.
Indeed, if multiple sources are not independent, they are essentially seen as a single independent source, and are no more persuasive than a single source~\cite{Harkins1987}.
Take for example information pertaining to health: if four friends inform us that a diet works well, this is more credible when these friends do not know each other and reached this conclusion independently from one another than when they all know each other (e.g. from a diet club), and have influenced each other in believing that the diet works.

We analyse a network of over $50\,000$ elites coupled with donation data to investigate the dynamics of this complex contagion.
Overall, the elite studied here follows the classical definition of the ``power elite'' as conceptualised by Mills~\cite{Mills1956}: they are mostly people in positions of power, such as politicians, business leaders, lobbyists and top bureaucrats, although there are also some exceptions, such as academics and public intellectuals, but they are relatively few.
We examine whether contagion is driven by cohesive reinforcement or independent reinforcement.
We further study cross-cutting exposure~\cite{mutz_consequences_2002,Bakshy2015}: how exposure to Democratic donors affects donations to a Republican candidate and vice versa. 
Finally, we examine whether the micro level effect can be extrapolated to the macro level by using the network structure to predict the total amount of money raised.
We report results for the presidential election cycle of 2008 (January 1, 2007 to December 31, 2008) for the Republican and Democratic candidate in the main paper.
Results for other campaigns and for donations to the Democratic and Republican National Committees can be found in Appendix~\ref{sec:detailed_results}.

\section{Results}

\begin{figure*}[tb]
  \centering
  \includegraphics{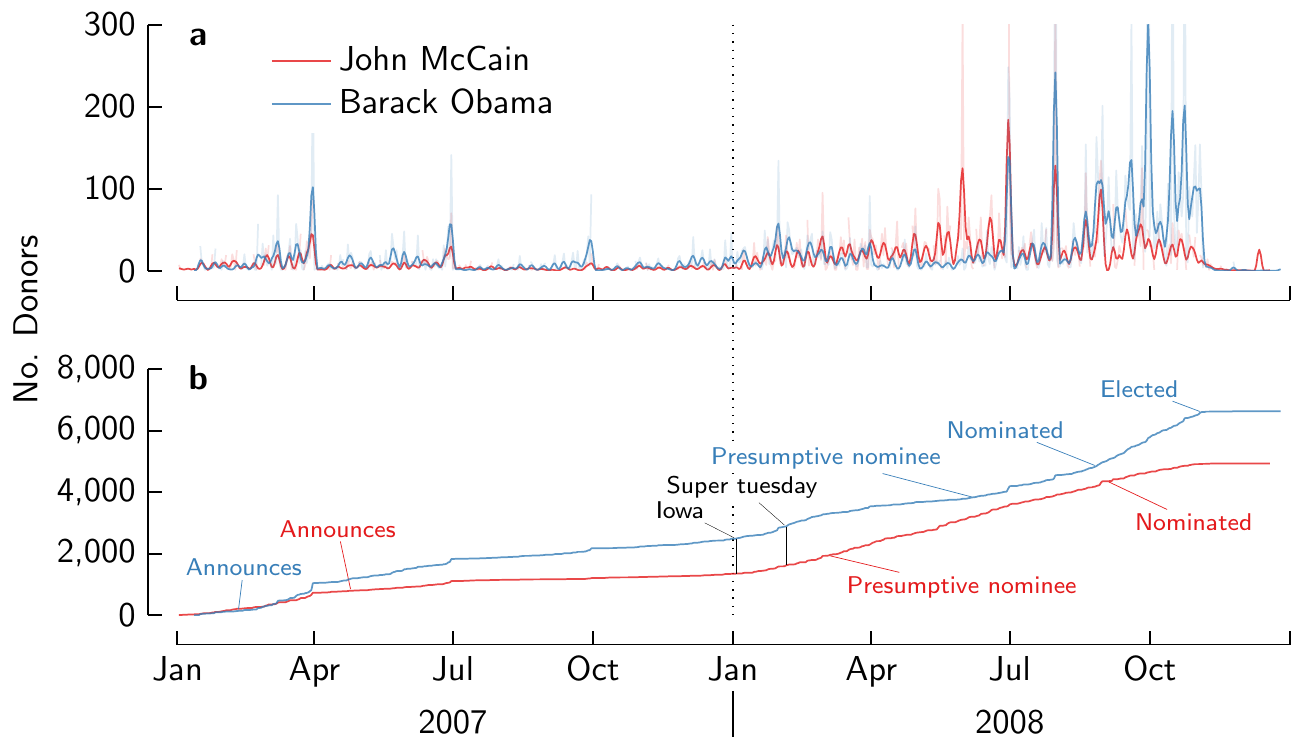}
  \caption{
    \textbf{Donor dynamics.}
    Daily donor dynamics (a) are affected by FEC deadlines (raw data is transparent, smoothed data solid).
    The cumulative number of donors (b) shows the overall growth.
  }
  \label{fig:donation_dynamics}
\end{figure*}

We use data gathered by \href{http://www.littlesis.org}{LittleSis}, a website that tracks US elites, for constructing a network between elites (see Appendix~\ref{sec:materials_and_methods}).
We construct the elite network based on direct family, professional and social relations (which constitute only a small fraction of relations), and indirectly on having worked at the same firm, being alumni of the same college or university, or being a member of the same organization or club.
As such, it is a multiplex network, which may have effects on the interplay between structure~\cite{Battiston2014,Omodei2015} and dynamics~\cite{Sole-Ribalta2015}.
As we will show later, the multiplexity indeed has an impact on the way donations diffuse.
The LittleSis database also records who donated how much to whom at what time. 
We study how the probability to donate depends on donors in the (network) neighbourhood.
In complex contagions, the probability to donate depends on the number of donors in the network neighbourhood which we call the \emph{donor degree}.
If contagion is reinforced through cohesion, only donors who are part of the same community are relevant.
In particular, the maximum number of donors who come from the same community should be pertinent, which we call the \emph{common community donor degree}.
If contagion takes places through independent reinforcement, only independent exposures are considered.
We operationalise this by counting the number of communities from which at least one member has donated, which we call the \emph{donor communities}.
Alternatively, we count the types of sources that have donated (family, professional, social, colleague, fellow alumni, fellow (club)member), which we call \emph{source diversity}.
We discount any effect of multiple exposures from similar sources.
Notice this is a measure of multiplex exposure: we count the number of layers on which a node is exposed.
These measures are illustrated in Fig.~\ref{fig:network_example}, and are defined formally in Appendix~\ref{sec:materials_and_methods}.
We take a quarter as a unit of time, since the daily dynamics exhibit clear dependencies on the (quarterly or monthly) FEC report deadlines (Fig.~\ref{fig:donation_dynamics}), as found previously~\cite{christenson_riding_2011}. 
We pool data for all quarters.

\begin{figure}[tb]
  \centering
  \includegraphics{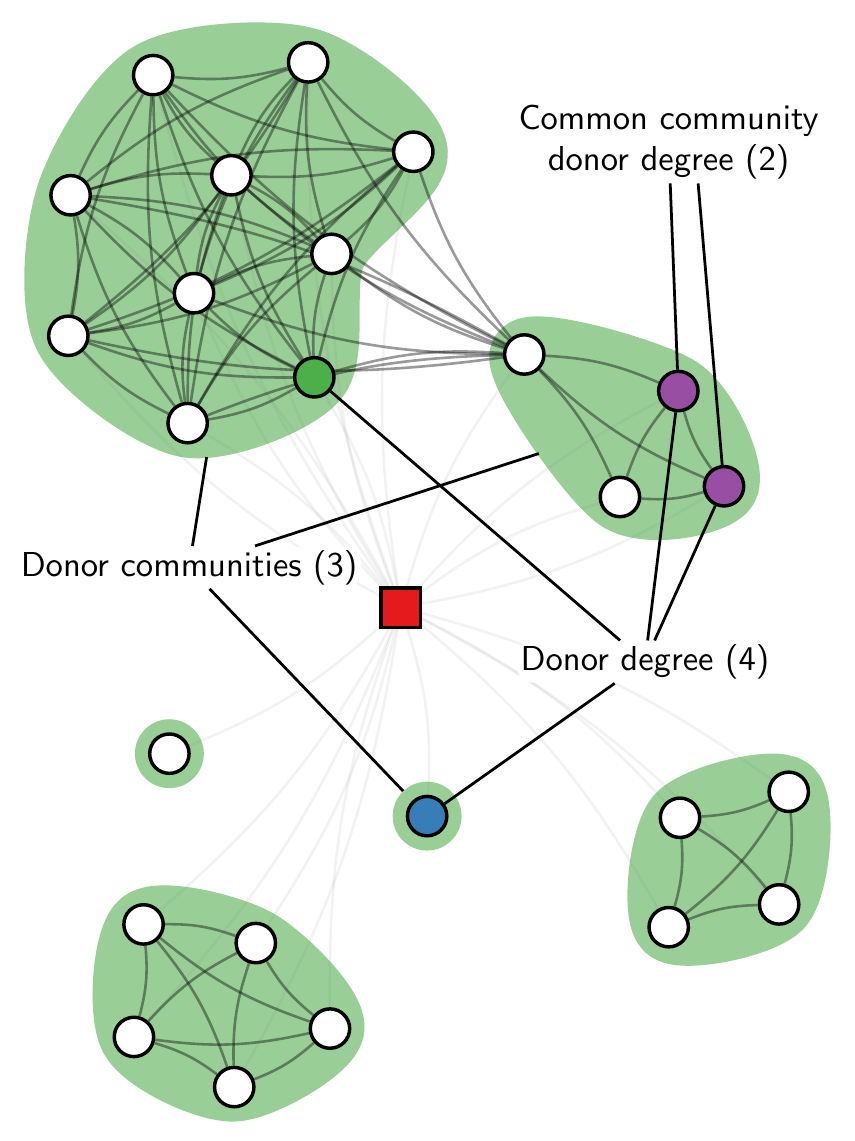}
  \caption{
  \textbf{Network neighbourhood example.}
  The focal node in the centre (square) is related to all the other nodes (shown in light lines), who may be connected to each other (shown in dark lines). 
  The neighbourhood can be divided into communities (shown in shades of green). 
  Filled nodes have donated, and the focal node is surrounded by four donors, who are in three different communities, with a maximum of two nodes from the same community.
  The central question is whether contagion of donation is driven by cohesive reinforcement---in which case the common community donor degree should have a large effect---or by independent reinforcement---in which case the donor communities should have a large effect.}
  \label{fig:network_example}
\end{figure}

\begin{figure*}[tb]
  \centering
  \includegraphics{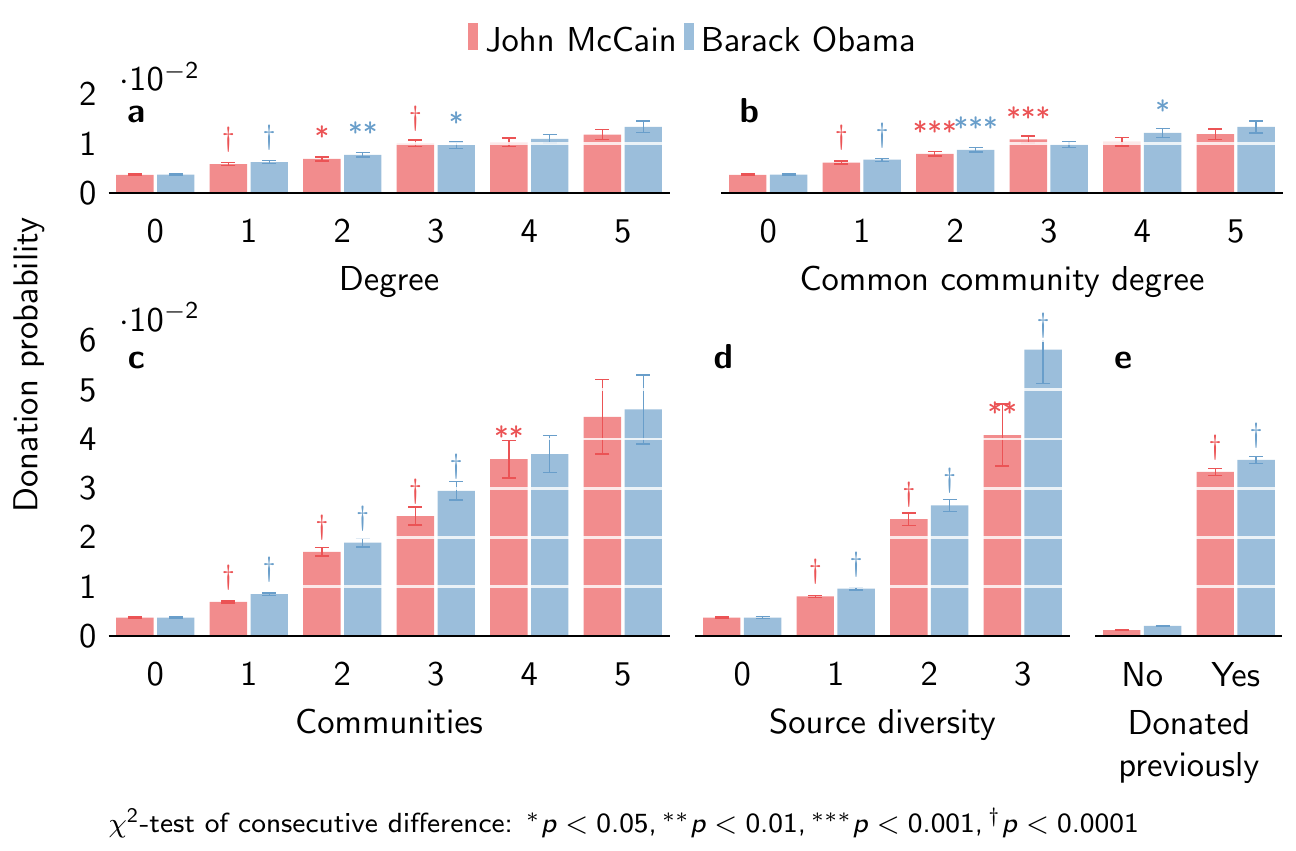}
  \caption{
    \textbf{Contagion effects.}
    The probability to donate based on (a) donor degree, (b) common community donor degree, (c) donor communities, (d) source diversity and (e) previous donation.
  }
  \label{fig:donation_prob}
\end{figure*}
\subsection{Contagion of donation}

The main question we address in this section is whether contagion is driven by cohesive reinforcement or by independent reinforcement.
We first analyse whether donation is more likely when exposed to donors by studying the effect of the \emph{donor degree}.
We then compare this to the effect of the number of independent donors.
This entails comparing the effect of the \emph{(common community) donor degree} to the effect of \emph{donor communities} and \emph{source diversity}.
We first study this question by reporting the immediate effects of exposure.
However, these effects may be biased due to homophily---as we will explain later---and we therefore check our findings by controlling for previous donations.
Finally, we corroborate our results using logistic regression in order to consider all factors simultaneously.

The probability to donate in the next quarter clearly depends on the donor degree (Fig.~\ref{fig:donation_prob}(a)).
A person exposed to a single donor is $1.7$ times more likely to donate than a person not exposed to any donor ($\chi^2 = 71.6$, $p = 2.68 \times 10^{-17}$ for Democrats, results for Republicans are comparable).
The marginal effect of exposure to more donors decreases: exposure to two donors makes donation $1.23$ times more likely than exposure to a single donor ($\chi^2 = 7.2$, $p = 0.0072$); exposure to further donors shows diminishing returns.
There is hence a clear effect of increasing exposure to donors.
The marginal effect is slightly stronger in the common community donor degree (Fig.~\ref{fig:donation_prob}(b)), where exposure to a single donor makes donation $1.8$ times more likely than exposure to no donors ($\chi^2 = 101.8$, $p = 6.07 \times 10^{-24}$).
People exposed to two donors from the same community are in turn $1.3$ times as likely to donate than people exposed to a single donor ($\chi^2 = 13.2$, $p = \num{0.00029}$).
The effect size of the common community donor degree is only slightly stronger, making the case for cohesive reinforcement not very strong.

The measures for independent reinforcement show much stronger effects.
Exposure to a single donor community makes donation $2.27$ times more likely ($\chi^2 = 320.5$, $p = 1.15 \times 10^{-71}$). 
An additional donor community makes donation $2.23$ times more likely again ($\chi^2 = 244.9$, $p = 3.45 \times 10^{-55}$, Fig.~\ref{fig:donation_prob}(c)).
The marginal effect diminishes: three donor communities make donation only $1.56$ times more likely ($\chi^2 = 32.3$, $p = 1.32 \times 10^{-8}$), still high compared to the effect of donor degree.
Overall, the probability to donate after exposure to two donor communities is of the same order as the maximum effect of donor degree (either overall or common community).
Indeed, when controlling for donor communities, donor degree shows no significant effect.
Independence measured in terms of source diversity shows even stronger effects (Fig.~\ref{fig:donation_prob}(d)).
Exposure to a single type of source increases the likelihood of donation $2.56$ times ($\chi^2 = 467.2$, $p = 1.31 \times 10^{-103}$).
A second type of source increases the likelihood of donation a further $2.77$ times ($\chi^2 = \num{404.2}$, $p = 6.57 \times 10^{-90}$).
While exposure from a third type of source increases the likelihood of donation again $2.19$ times ($\chi^2 = \num{37.8}$, $p = 7.92 \times 10^{-10}$), the effect weakens thereafter.
In short, there is evidence of independent reinforcement.

One of the notorious problems of studying social contagion is the problem of homophily~\cite{Aral2009}.
People with similar preferences are more likely to be connected~\cite{McPherson2001}, so that exposure to donors does not necessarily causally affect donation, but may simply reflect an underlying similarity in preferences.
Such homophilic links may have consequences for how phenomena diffuse over networks~\cite{Centola2011,Alstott2014,Anderson2015,DelVicario2016}.
Homophily can also create a selection bias (treatment bias) as exposure to donors also depends on underlying preferences, meaning that donors are more likely to be exposed to other donors~\cite{Aral2009}.
This problem has previously been addressed by resorting to propensity matching methods~\cite{Aral2009}.
While it is impossible to completely resolve the confounding of homophily and social contagion in observational studies~\cite{Shalizi2011}, contagious effects have also been documented in experimental studies~\cite{Muchnik2013}.
The single best indicator of the underlying preference for donating to a particular campaign is whether that person has previously donated to other Democratic or Republican candidates (Fig.~\ref{fig:donation_prob}(e)).
Note that campaigns often use previous donations to target likely donors~\cite{Hassell2013}, so that controlling for previous donations effectively also controls for selective targeting by campaigns.
Previous donation makes it more than $17$ times more likely to donate again ($\chi^2 = \num{7801}$, $p \approx 0$).
We use this measure to control for homophily.
We cannot rule out other hidden homophily that might explain any remaining network effects~\cite{Shalizi2011}.
While other relevant co-variates are available in the data (e.g. date of birth, gender, net worth), they have not been as extensively coded, limiting their usefulness as control variables.
Nevertheless, observing an identical phenomenon (donation to a political candidate) is a strong indicator that should account better than any other measure for the effects of homophily on donation.
If we still see an effect of exposure to other donors after controlling for previous donations---which is such a strong predictor---then this strengthens the hypothesis of social contagion.

\begin{figure*}[tb]
  \centering
  \includegraphics{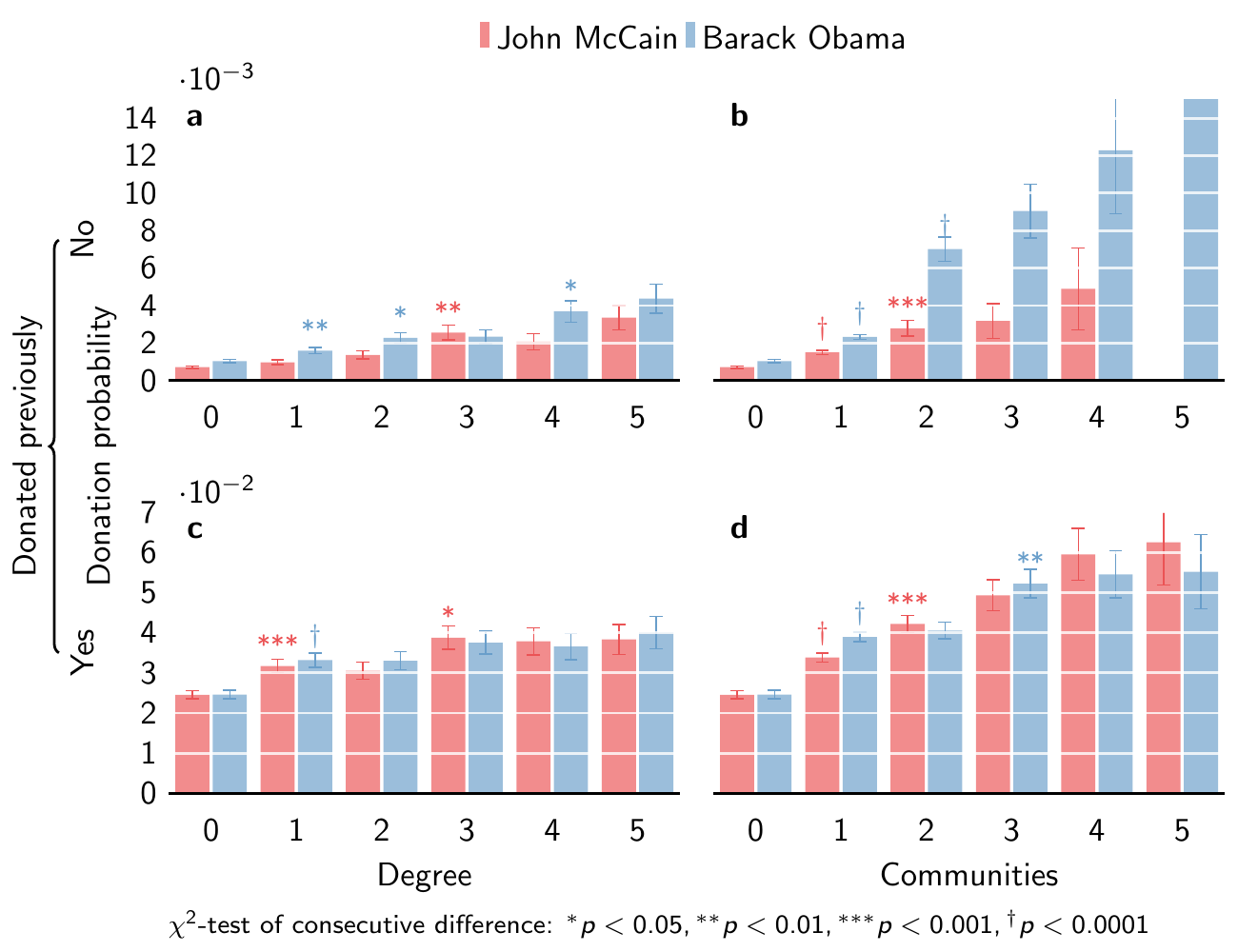}
  \caption{
    \textbf{Conditional donation probability.}
    Donation conditional on previous donation differs for (a,c) donor degree and (b,d) donor communities, with the effects for new donors being larger.
  }
  \label{fig:donation_conditional}
\end{figure*}

We separate the effects of donor degree and donor communities by previous donation (Fig.~\ref{fig:donation_conditional}). 
Overall, donation is much more likely from old donors (those who have previously donated) than from new donors (those who have not), as said earlier.
Exposure to a single donor increases the likelihood $1.53$ times for new donors ($\chi^2 = 10.3$, $p=0.0013$), and only $1.35$ times for old donors ($\chi^2 = 18.4$, $p = 1.76 \times 10^{-5}$).
An additional exposure increases the likelihood of donation a further $1.43$ times for new donors ($\chi^2 = 4.65$, $p = 0.031$); for old donors, additional exposure does not increase the probability of donation further.
The effect is clearly stronger for new donors than for old donors.
The effect of independent reinforcement is again much stronger, although a similar difference between old and new donors is apparent.
For new donors, the likelihood of donation is $2.24$ times higher for a single community exposure ($\chi^2 = 73.4$, $p = 1.04 \times 10^{-17}$), while for old donors this is only $1.58$ ($\chi^2 = 76.05$, $2.77 \times 10^{-18}$).
An additional community exposure for new donors makes donation $3.01$ times more likely ($\chi^2 = 111.9$, $p = 3.80 \times 10^{-26}$), while for old donors donation is only $1.04$ times more likely ($\chi^2 = 0.42$, $p = 0.52$).

\begin{figure*}[tb]
  \centering
  \includegraphics{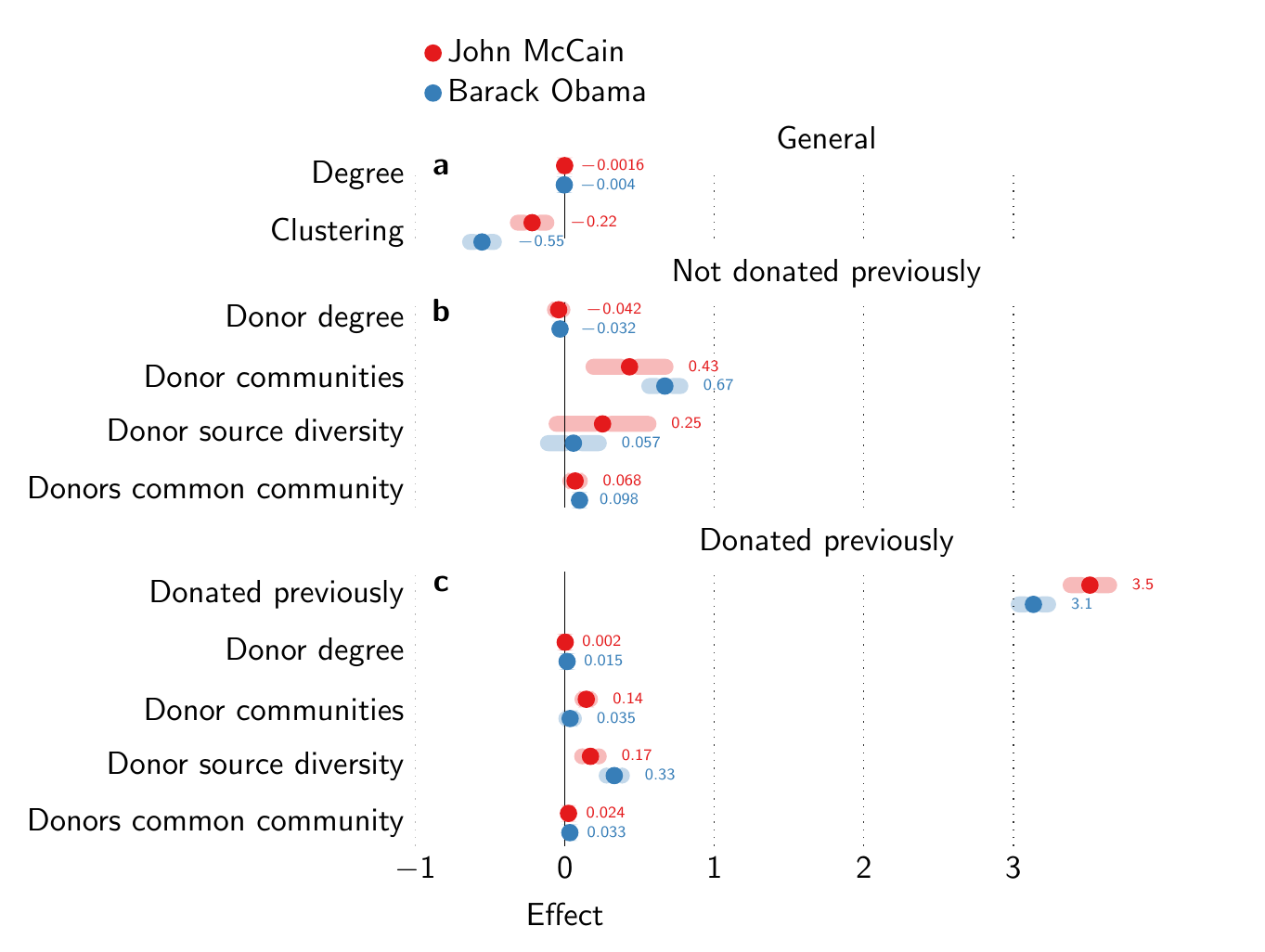}
  \caption{
      \textbf{Logistic regression results.}
      Magnitude of (a) general effects, (b) network effects for new donors, and (c) network effects for old donors.
      Error bars show 95\% confidence intervals for the coefficients.}
  \label{fig:regression}
\end{figure*}

Logistic regression confirms that donor communities have the strongest effect, and that the effects are stronger for new donors than for old donors (Fig.~\ref{fig:regression}, Tables \ref{SI_tab:regression_candidates_D} and \ref{SI_tab:regression_candidates_R}).
We also control for overall degree (i.e. not only including donors), which has a slightly negative effect ($\theta = -0.0040$, $t = \num{-11.7}$, $p = 1.40 \times 10^{-31}$) and clustering of a node, which has a somewhat stronger negative effect ($\theta = -0.55$, $t = -7.95$, $p = 1.93 \times 10^{-15}$).
Hence, hubs and brokers are less likely to donate overall.
When controlled for all other variables, donor degree actually slightly decreases the chances of donation for a new donor ($\theta = \num{-0.032}$, $t = \num{-2.00}$, $p = \num{0.046}$), but the effect is positive for old donors ($\theta = 0.015$, $t = 2.12$, $p = 0.034$).
Common community donor degree does have a positive effect on new donors ($\theta = 0.098$, $t = 5.40$,
$p = 6.63 \times 10^{-8}$) and old donors ($\theta = 0.033$, $t = 4.04$, $p = 5.41 \times 10^{-5}$), multiplying the odds of donation about $1.10$ times and $1.03$ times for each additional common community donor degree.
Donor communities have a much stronger effect on new donors ($\theta = 0.67$, $t = 8.55$, $p = 1.24 \times 10^{-17}$), increasing the odds of donation $1.95$ times for each additional community, but have no significant effect on old donors ($p = 0.38$).
Finally, source diversity has no significant effect for new donors ($p = 0.61$), but does have a clear positive effect on old donors ($\theta = 0.33$, $t = 6.19$, $p = 6.03 \times 10^{-10}$), increasing the odds of donation $1.39$ times.
The strongest effect clearly remains previous donation ($\theta = 3.13$, $t = 41.05$, $p \approx 0$), increasing the odds $23$ times.

In conclusion, we find that the contagion of donation is mostly driven by independent reinforcement contagion.
For new donors, the number of communities has a much larger effect than donor degree (either overall or community degree).
Donations in their neighbourhood increase the likelihood of donation, especially when the previous donors come from different social groups.
For old donors, the effects are typically much weaker or even insignificant.
Their decision to donate is largely independent of donations in their neighbourhood.
The presidential election campaign of 2008 discussed here shows the clearest effects, perhaps related to the unexpected success of Barack Obama.
Results for other election cycles of 2000, 2004 and 2012 and for party donations---to the Democratic National Committee (DNC) or Republican National Committee (RNC)---are qualitatively similar (see Figs. \ref{SI_fig:presidential_2000}--\ref{SI_fig:party_2012} and Tables \ref{SI_tab:regression_candidates_D}--\ref{SI_tab:regression_cross_party_R} in Appendix~\ref{sec:detailed_results}).

\subsection*{Cross-cutting exposure}

\begin{figure*}[tb]
  \centering
  \includegraphics{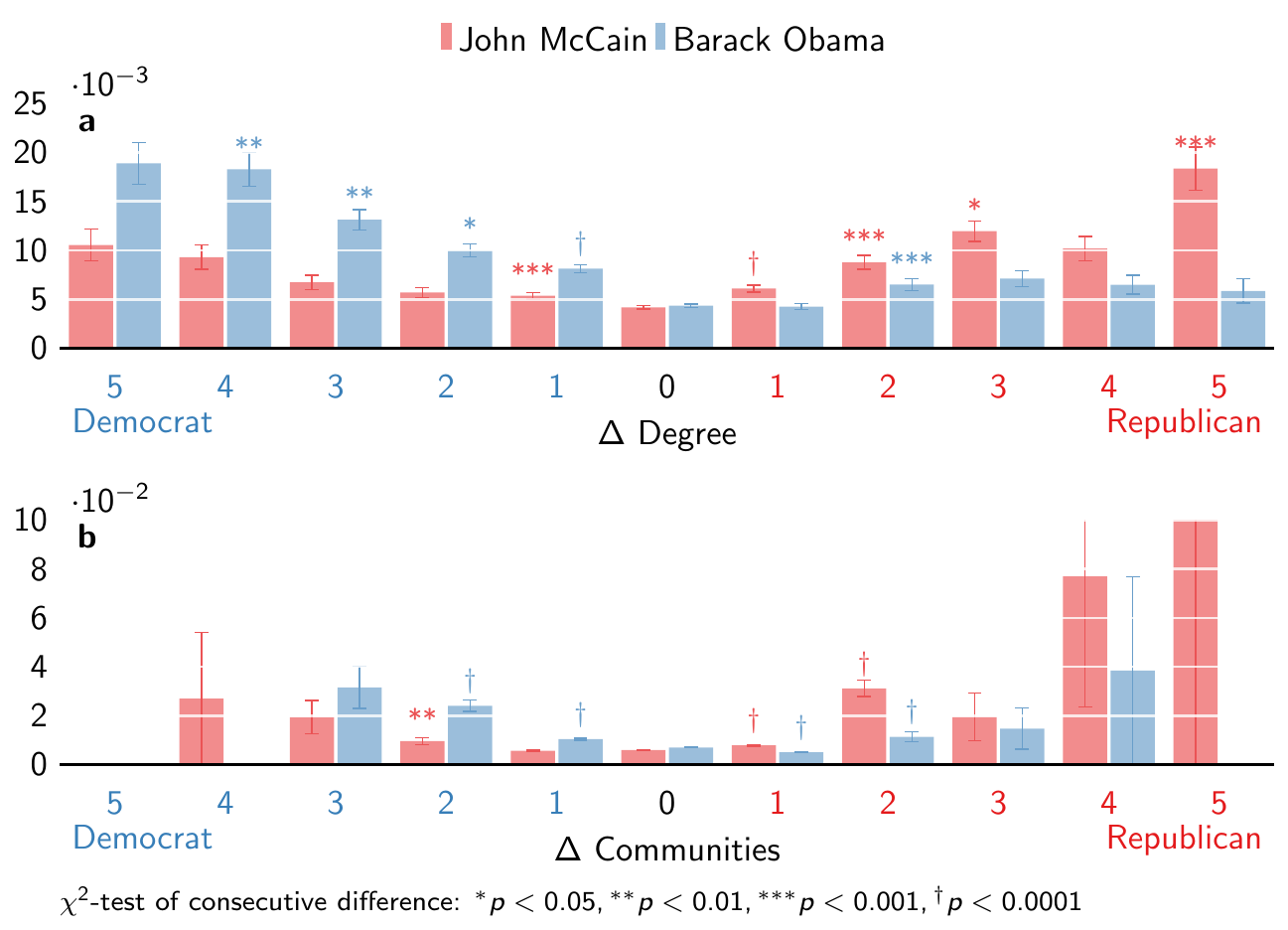}
  \caption{
    \textbf{Cross-cutting donation probability.}
    Cross-cutting donation probability by (a) degree and (b) communities shows that exposure to one side can increase donations to the other side.
  }
  \label{fig:donation_cross_cutting}
\end{figure*}

In the previous section, we found that exposure to Democratic donors makes donation to the Democratic candidate much more likely (and similarly so for Republicans), especially for new donors that are exposed to multiple independent donors.
It is possible that donation is also affected by the amount of support for the opposing party.
For example, if somebody is equally much exposed to Democratic and Republican donors, it might put him or her in a rather difficult position: should (s)he support the Democratic or the Republican candidate?
One possibility is that (s)he donates to neither, to prevent possible conflicts, which was found to be the case for voting~\cite{mutz_consequences_2002}.
In this section we analyse the effects of this so-called cross-cutting exposure to the ``other'' party (i.e. the effect of Democratic exposure on Republican donations and vice-versa).

Cross-cutting exposure has a counter-intuitive effect.
If there are relatively more Democrats (taking the difference between the number of Democrats and Republicans), this not only increases the likelihood of donations to the Democratic candidate, but also to the Republican candidate (Fig.~\ref{fig:donation_cross_cutting}).
Exposure to one more Democrat than Republican donor increases the likelihood of donations to Republicans $1.28$ times ($\chi^2 = 11.2$, $p = 8.39 \times 10^{-4}$).
While Republican donations continue to increase with further exposure to Democratic donors, none of the remaining consecutive differences are significant.
The effect of relatively more exposure to Republican donors remains greater, with a single exposure making Republican donations $1.46$ times more likely ($\chi^2 = 28.2$, $p = 1.09 \times 10^{-7}$).
The effect seems less pronounced for the relative number of communities in this analysis.

\begin{figure*}[tb]
  \centering
  \includegraphics{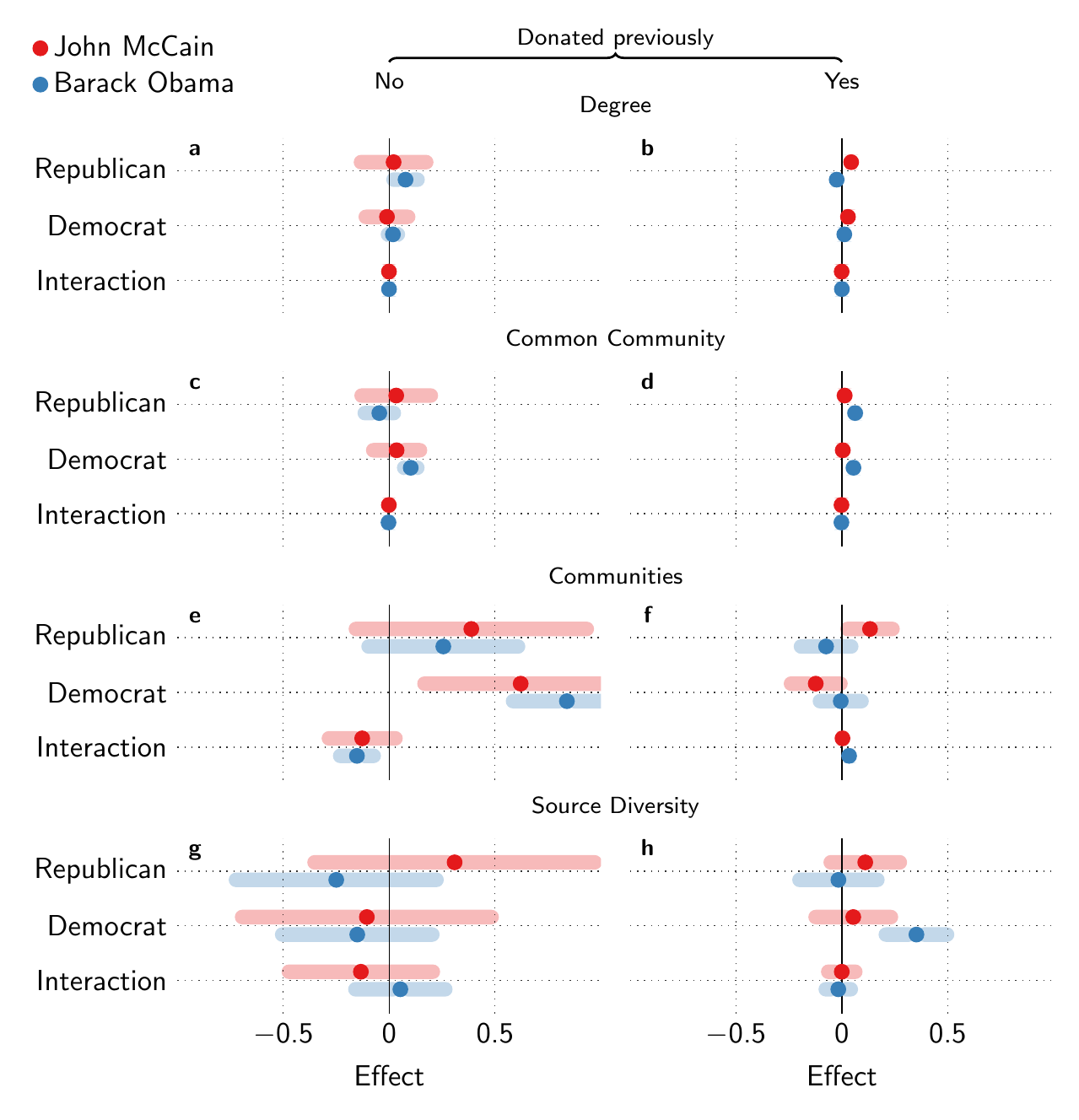}
  \caption{
  \textbf{Logistic regression results for cross-cutting effects.}
    Effect sizes for cross-cutting exposure distinguished by old/new donors for (a)--(b) donor degree, (c)--(d) common community donor degree, (e)--(f) donor communities, and (g)--(h) source diversity.
    Error bars show 95\% confidence intervals for the coefficients.
    }
  \label{fig:regression_cross}
\end{figure*}

However, logistic regression shows that exposure to Democratic donor communities increases the odds of donation to a Republican candidate ($\theta = 0.62$, $t = 2.73$, $p = 0.0063$), and that none of the other Democratic exposure measures are significant (Fig.~\ref{fig:regression_cross}).
In fact, the effect of Democratic donor communities is higher than Republican donor communities for Republican donations ($\theta = 0.39$, $t = 1.47$, $p=0.14$).
The inverse is not significant---exposure to Republican communities does not increase donation to the Democratic candidate significantly.
For Democratic donations, the interaction of exposure to Democratic and Republican community donors is $-0.15$ ($t = -2.10$, $p = 0.036$), but this is not significant for Republican donations ($p = 0.13$).
Similarly, the interaction of Democratic and Republican common community donor degree has a slight negative effect on Democratic donations ($\theta = -0.0031$, $t = -2.51$ $p = 0.012$), which is again not significant for Republican donations.
This implies that when simultaneously exposed to both Republicans and Democrats, Democratic donations may become less likely, in line with previous results~\cite{mutz_consequences_2002}, but this is not the case for Republican donations.

Interestingly, effects of cross-cutting exposure vary per campaign.
For example, while Republican donations were more likely after exposure to Democratic donors in the 2008 election, the reverse was true in 2004: exposure to Republican donors triggered Democratic donations (Figs. \ref{SI_fig:presidential_2004} and \ref{SI_fig:party_2004}).
Another example, in the 2012 election, any community exposure (either to Democratic or Republican donors) led to Republican party donations and neither had an impact on Democratic donations (Figs. \ref{SI_fig:presidential_2012} and~\ref{SI_fig:party_2012}).
See Appendix~\ref{sec:detailed_results} for further details on cross-cutting effects for other election cycles.
This may suggest that such cross-cutting exposure may be triggered by specific dynamics of some campaigns.
We briefly discuss this further in Section~\ref{sec:discussion}.

\subsection{Overall campaign}

We have seen that, at a micro level, when somebody is exposed to multiple independent donors, (s)he is more likely to donate.
This micro effect can possibly also have network wide repercussions at a macro level.
We may expect that the total extent of the diffusion is greater if people from different communities have donated than if equally many people have donated from the same community.
We here briefly investigate whether this extrapolation of the micro effect to the macro level holds.
 
\begin{figure*}[tb]
  \centering
  \includegraphics{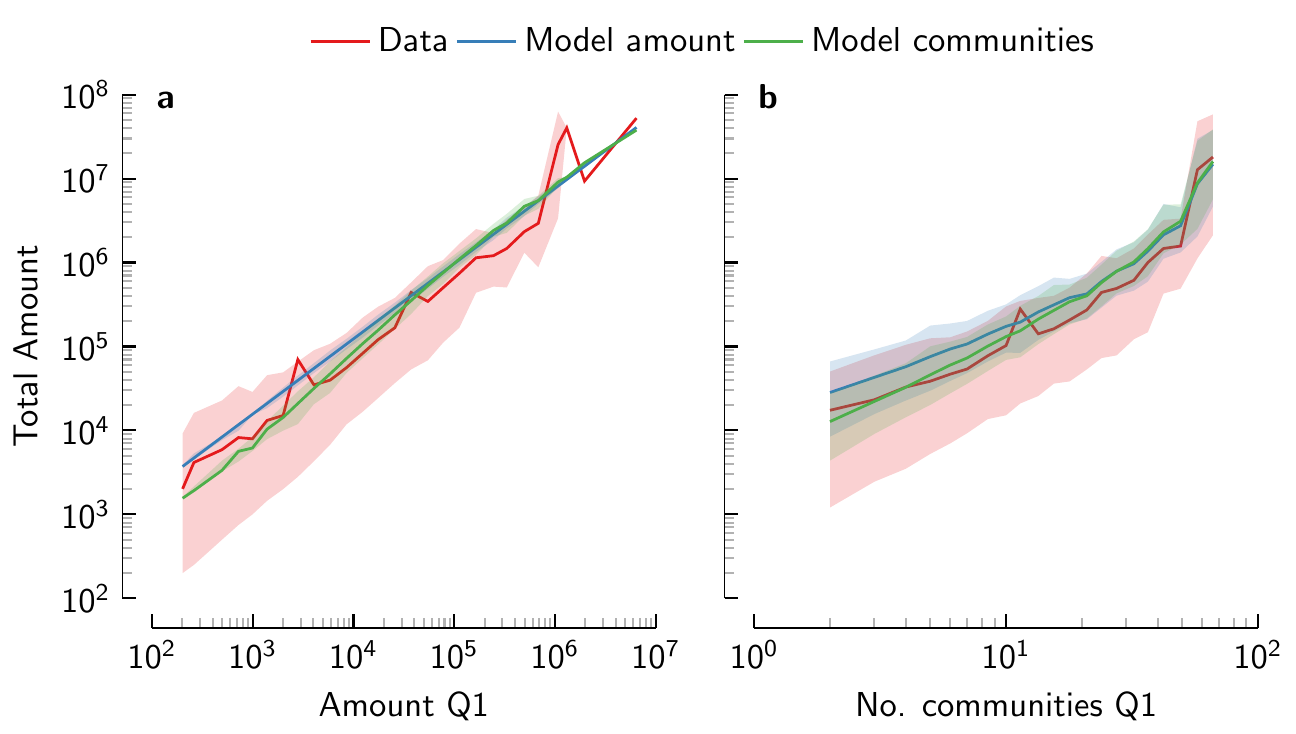}
  \caption{
    \textbf{Predicting total campaign contributions.}
    We predict the total amount donated throughout the campaign based on donations in the first quarter only. Line shows average and the shade shows the 5\% and 95\% percentiles.}
  \label{fig:donation_total_amount}
\end{figure*}

Following the extrapolation, the number of communities that donate at some point should be indicative of the amount of money raised afterwards.
To test this, we try to predict the total amount of money $w_i$ raised in a certain election cycle for any candidate (including senatorial, congressional and presidential elections) based on the first quarter (Fig.~\ref{fig:donation_total_amount})
The total amount of money raised throughout the campaign largely depends on the amount of money raised in the first quarter $w_i(1)$, and we find that $\hat{w}_i = \alpha w_i(1)^\beta$ predicts $w_i$ fairly well with $\alpha = 32.1 \pm 4.2$ and $\beta= 0.90 \pm 0.0086$.
Taking into account the number of communities that have donated in the first quarter $c_i(1)$ as $\hat{w}_i = \alpha w_i(1)^\beta c_i(1)^\gamma$ with $\alpha = 21.6 \pm 3.7$, $\beta = 0.81 \pm 0.015$, and $\gamma = 0.42 \pm 0.070$ slightly improves the fit, and is clearly favoured ($\Delta \AIC = 30.32$).
The number of donors itself has no significant effect and actually degrades the fit, clearly favouring the model using the number of communities.
Both $\beta$ and $\gamma < 1$, so that additional donations and independent donors yield diminishing returns.
Every doubling of the amount raised in the first quarter multiplies the total amount by roughly $1.75$.
Doubling the initial number of communities multiplies the total amount by about $1.34$, while doubling the initial number of donors has no significant effect.
This small exercise demonstrates that the effect of independent donors not only holds at the micro level, but can also be extrapolated to the macro level.

\section{Discussion}
\label{sec:discussion}

We find that political campaign donations among elite are socially contagious.
Being exposed to other donors increases chances of donation significantly, also after controlling for previous donations.
Contagion is especially likely after multiple exposures from different communities or from different types of sources (e.g. family, friends, business partners).
This supports the idea of independent reinforcement of complex social contagions.

Our results show that having a multiplex view of a network may be important for understanding the diffusion process.
In particular, the perceived independence of people may be difficult to asses without distinguishing different types of links.
Some of this information may also be contained in the ``flat'' network (where we disregard the type of a link) depending on how the structure is correlated between different multiplex layers.
If there is a positive correlation between two layers, a single community in the flat network may consist of multiple types of links, while if this correlation is negative, different types of links are likely to be contained within different communities.
The latter seems to be the case for the LittleSis data, where only about $3\,000$ links out of more than $1.6$ million links are of more than one type, explaining the congruency between community degree and different types of sources.

We theorize that people infer population-wide behaviour by observing their local networks.
Such local observations are necessarily biased, as they are influenced by homophily and other selection effects.
One reasonable heuristic for trying to surmount such bias would be to only use observations that are as independent as possible.
Whenever ego knows two observations are not independent, (s)he only counts them as one~\cite{Harkins1987}; (s)he discards redundant observations.
Alternatively, if two persons of the same background show the same behaviour, people attribute it to those particular characteristics or group membership~\cite{Wilder}, whereas if two people act the same but have very different backgrounds, people attribute the behaviour to its popularity.
Stated differently: people infer that the TV show \emph{The Big Bang Theory} is widely popular when not only their geek friends watch it, but when friends from all backgrounds watch it.
It would be interesting to further examine this hypothesis of attribution bias~\cite{Heider1958}.
This is consistent with the idea of the ``majority illusion'', generalizing local observations to population-wide estimates leading to a locality bias~\cite{Lerman2015}.

It is well known that viability of candidates plays a major role in campaign donations~\cite{dowdle_viability_2009,meirowitz_contributions_2005,hassell_looking_2011}.
Campaign donations play an especially important role during primary nominations~\cite{steger_primary_2000,norrander_attrition_2006}, as do (party) elites more generally~\cite{anderson_which_2013}.
When a candidate's success seems out of reach, people are reluctant to donate.
After all, supporting a (guaranteed) losing candidate simply squanders money, which could be allocated to a more viable candidate.
Using independent observations to infer the viability of candidates seems a good heuristic, especially early on in the campaign when candidates are not yet widely known. 
We theorize that independent reinforcement is especially relevant for campaign donations to assess viability. 

More generally speaking, we hypothesize that complex contagion with independent reinforcement is especially likely with population-wide network externalities---if the value of the behaviour depends on the number of people showing such behaviour~\cite{EasleyKleinberg2010}.
A previous study of contagion in signing up to Facebook reported similar results to our own~\cite{Ugander2012}; a clear case of behaviour with network externalities---using a social network site is only useful if enough other people use it.
Cases where network externalities are present are abundant, ranging from communication devices and services to file formats and technical standards~\cite{Katz}.
In such situations we expect complex contagions to be driven by independent reinforcement.

However, not all network externalities necessarily lead to independent reinforcement.
For example, in the presence of a public goods dilemma or a collective action problem, we may expect cohesive reinforcement rather than independent reinforcement.
If two alters are unrelated (such that ego is the broker between these two alters), this may make coordination between the three more difficult; when the alters do know each other, it may facilitate communication and thereby agreement on a common course of action.
Especially when the stakes are high, failing to coordinate an action can be costly, thereby making cohesive reinforcement more likely.
While cascades of cooperation have been observed to take place in experimental studies of public goods~\cite{Fowler2010,Rand2014,Tsvetkova2014,Suri2011}, it is an open question whether such effects are driven by cohesive or independent reinforcement contagion.

Alternatively, if a network externality depends not on the number of people, but on the connections between them, we may expect cohesive reinforcement.
Take the mundane example of hanging out at a bar with friends.
The joy of such an activity depends not on the number of friends itself, but on whether those friends also know each other and enjoy spending time with each other.
Similarly, team performance may not depend so much on the number of people, but on how they are connected~\cite{DeMontjoye2014,Balkundi2006}, favouring cohesive reinforcement. 

Inferring population-wide estimates from local observations may not only have positive contagious effects.
Knowing that sufficient people are already contributing to a public good may make it less likely for people to contribute themselves~\cite{Tsvetkova2014,Suri2011}.
When estimates of contribution to a public good are derived from local estimates, we may see a negative contagion effect.
In our case, this could explain the counter-intuitive finding that exposure to one side can trigger donations to the other side.
Here, people could deem it necessary to rally support for their candidate of choice because they observe too much support for the opponent.
In the 2004 campaign, when George W. Bush was seeking re-election, potential Democratic donors may have been reacting to exposure to donations to Bush to oppose his candidacy.
Similarly in the 2008 campaign, potential Republican donors may have wanted to oppose Obama's election.
More generally speaking, people may react to their local estimate of population-wide behaviour rather than the actual population-wide behaviour.

Our findings may also have implications for recommendation systems in online social networks~\cite{Konstas2009}. 
It may be more relevant what story or product is liked or bought by many independent friends than by friends from the same social group. 
This is also what the results of the study on joining Facebook suggests~\cite{Ugander2012}: the adoption of the service depends on the number of independent friends that have joined Facebook.
However, this may depend on the type of product which is recommended.

The two different modes of complex contagion also affect the speed of the spreading process on networks~\cite{Centola2010,Nematzadeh2014}.
Independent reinforcement has a relatively slow and steady rate of diffusion in clustered networks, as additional contagions in the local neighbourhood are redundant and do not reinforce contagion.
In contrast to cohesive reinforcement, independent reinforcement may permeate group boundaries.
Depending on the type of contagion and the type of network, we may expect different dynamics, which should be explored further.

Finally, as we stated at the outset, money is vital to any political campaign in the US.
Raising money is a crucial short-term goal in order to further the end goal of getting elected.
Our findings suggest that appealing to constituencies of diverse backgrounds may actually aid in diffusing support through networks. 
Moreover, the number of communities that have donated is significantly predictive of total fund-raising capabilities, whereas the number of donors is not.
The number of communities was also found to be predictive of the virality of online memes~\cite{Weng2013}.
While this is congruent with the idea that independent reinforcement takes place on the network, it could also indicate the candidate's more widespread appeal---both interpretations support the strategy of targeting people in communities that have not yet donated. 
Doubling the amount of money raised but only targeting the same communities only multiplies the total amount by about $1.75$, while doubling the number of communities at the same time more than doubles the amount of money, resulting in a super-linear scaling.
Contagious effects may multiply the efforts of fund-raising, and they should be taken into account.
This suggests a (perhaps counter-intuitive) change to fund-raising strategy, suggested earlier in a study on soliciting donations~\cite{lim_mobilizing_2010}: it may be more effective to target people who are relatively difficult to recruit, rather than the easy picks.
Although targeting likely donors may seem to have greater direct effects with relatively more people assenting to the request---targeting various groups may have a larger effect overall because donations are more likely to spread, even though fewer may immediately assent to the request.
We consider this a hopeful sign, suggesting that rather than addressing narrow interests and petty concerns, politicians should appeal to the general population and the greater good.

\begin{acknowledgments}
This work is supported by the Royal Netherlands Academy of Arts and Sciences (KNAW) through its \href{http://www.ehumanities.nl/}{eHumanities project}. We are grateful to \href{http://www.littlesis.org}{LittleSis} for making its data available for analysis.
\end{acknowledgments}

%

\setcounter{figure}{0}
\setcounter{table}{0}
\renewcommand*{\theHfigure}{S\arabic{figure}} 
\renewcommand\thefigure{S\arabic{figure}}
\renewcommand*{\theHtable}{S\arabic{table}} 
\renewcommand\thetable{S\arabic{table}}

\appendix

\section{Materials and Methods}
\label{sec:materials_and_methods}

\subsection{LittleSis} 

\href{http://www.littlesis.org}{LittleSis} (as the opposite of Big Brother) is a website launched in 2009 to track US elites, including business leaders, lobbyists, politicians, bankers, top bureaucrats and others, totalling over $120\,000$ persons.
In addition, LittleSis provides data on over $40\,000$ organizations, including firms, labour unions, universities and private clubs.
Persons and organizations are tied to each other in several ways, detailing what positions people held in which organizations, where they were educated and of which organizations they are members.
The website also details family, social and other professional links between elites.
Last but not least, the website holds extensive records of who donated to whom, in the context of political campaigns, obtained from  the Federal Election Committee (FEC).
The LittleSis project is maintained by the Public Accountability Initiative, a non-profit, non-partisan organization.
It gathers this information from a variety of public sources, and brings it together in a single searchable interface, providing easy access to journalists and scholars trying to untangle the web of influence.
Alongside making this information publicly available through a search interface and a Web API, the LittleSis project renders a complete (SQL) database dump available upon request, which we use for our analysis.
Although all information is publicly available, we anonymised all data since identifying information plays no role in our analysis.

We constructed a network between all people included in the database.
We did so based on direct family, professional and social relations (which constitute only a small fraction of relations), and indirectly on having worked at the same firm, being alumni of the same college or university, or being a member of the same organization or club.
These indirect links only provide some indication of a possible connection between people, and not all connections are guaranteed to exist in reality.
Assuming that the number of connections people have to (former) colleagues, fellow alumni or fellow members is limited, the probability that two people actually have a connection generally decreases with the size of the organisation.
We therefore restrict these indirect links to organizations having fewer than 300 members in the LittleSis database.
To check for the robustness of our results with respect to this cut-off, we also run all analyses with a cut-off of 250 members and 500 members, and find that the results remain qualitatively the same.
Out of the nearly $120\,000$ people, $61\,196$ had at least one type of link, and we only used the largest connected component which includes $55\,435$ people.

The network is quite clustered (clustering coefficient 0.87), and has a quite high average degree of 58.5.
To get an idea of the type of links connecting members of the US elite in the LittleSis database, we remove specific types of links and look at the size of the remaining largest connected component.
Business relations hold most of the network together, resulting in a largest connected component of only a quarter after removing them.
This is not surprising, as business relations also comprise most of the links.
Family relations have a surprisingly large impact on connectivity compared to organization membership.
Although both membership and family reduce the largest component by about 4\%, the latter needs only 0.19\% of the links to do so, while the first needs more than 14\% to do the same, so that removing a family link reduces the largest component on average by 0.64 nodes.
Similarly, professional and social relations reduce the largest component on average by respectively 0.42 and 0.61 nodes per link.
The indirect relations (membership, education and business) remove only very few nodes per link. 
Understandable, given that all people associated with the same organization are connected to each other, thereby creating many redundant links.
The indirect links are essential in connecting the network.
Unfortunately, there is no way of telling whether two people connected through such an indirect link actually know each other, less so whether they have exchanged information regarding their donations.
The use of such indirect links is a limitation of the LittleSis data, and preferably our results should be checked against more accurate network data (which, unfortunately, is more difficult to gather).
Of course, even for direct links, we cannot ascertain whether people have actually exchanged information on their donations.
However, such (indirect) links indicate there is a certain probability for the flow of such information, and that hence exposure did take place.

\subsection{Modelling contagion}
For each donation, we know who donated to whom and when.
We use time to analyse how donations diffuse through the network $G = (V,E)$ with nodes $V$ and links $E$.
For each time step we study how the donation probability depends on donors in the (network) neighbourhood.
Formally speaking, let $Y_i(t)$ denote the fact that person $i$ has donated at time $t$ to a certain campaign. 
We then try to predict $Y_i(t)$ based on $Y_j(\tau)$ for $\tau < t$ for neighbours $j$ of $i$ (i.e. if $(ij) \in E$).
For convenience we set $X_i(t) = 1$ if $Y_i(\tau) = 1$ for any $\tau \leq t$ and $X_i(t) = 0$ otherwise, that is $X_i(t) = \max_{\tau \leq t} Y_i(\tau)$ indicates whether $i$ has donated before (or at) time $t$.
People may donate multiple times to the same campaign (once in the primary and once in the general election for example), but we only study contagion for the first donation as later donations are most likely caused by the first donation rather than through contagion. 
We model $\Pr(Y_i(t) = 1 \mid X_i(t - 1) = 0) = f(\theta, N_i, X(t - 1))$ where $N_i$ is the subgraph induced by the neighbours $\{ j | (ij) \in E\}$ and $\theta$ some parameters.

We use several characteristics of the neighbourhood graph $N_i$ for predicting donation.
We call the number of neighbours the \textit{degree} $k_i = |V(N_i)|$, and the density of the connections among the neighbours is called the clustering coefficient, which is $c_i = \frac{2 m_i}{k_i (k_i - 1)}$, where $m_i = |E(N_i)|$ is the number of links in the neighbourhood graph.
The nodes that have donated before or at time $t$ are denoted by $D(t) = \{ i \mid X_i(t) = 1 \}$.
The number of donating neighbours is the \textit{donor degree} $k_i^\text{donor} = |D(t) \cap N_i |$.
We denote by $\sigma_j$ the community of node $j$, detected for each neighbourhood graph $N_i$ separately.
That is, we detect communities for each $N_i$, and $\sigma_j$ is the community of node $j$ with respect to the neighbourhood graph $N_i$.
Detecting communities at the local level of the neighbourhood graphs is consistent with how they are perceived by ego.
We detect communities using the Louvain algorithm~\cite{Blondel2008} to optimize modularity~\cite{Newman2004Finding}.
Although modularity has difficulty detecting communities in large graphs~\cite{Fortunato:2007p183}, the neighbourhood graphs $N_i$ are generally quite small, so this should not be a problem.
Detecting communities at the overall network level $G$ does not yield the same results, and exposure measured in this way yields weaker effects.
We operationalise cohesive reinforcement as the number of donors that come from the same community.
We take the maximum number of donors who come from the same community, ignoring any exposure from other communities with fewer donors.
We call this the \textit{common community donor degree} which is $k_i^\text{ccd} = \max_c |\{j \mid \sigma_j = c  \text{~and~} j \in D(t) \cap N_i\}|$.
We operationalise independent reinforcement by considering the number of different communities that have donated; exposure from multiple donors from the same community is ignored.
We simply count the number of communities which have donors, which we call the \textit{donor communities}: $c_i^\text{donor} = |\{ \sigma_j \mid j \in D(t) \cap N_i\}|$, which is a proper set so that multiple $\sigma_j$ are only counted once.
Note that $k_i \geq k_i^\text{donor} \geq c_i^\text{donor}$ and $k_i^\text{donor} \geq k_i^\text{ccd}$ but that $c_i^\text{donor}$ and $k_i^\text{ccd}$ are not directly related, except that $k_i^\text{ccd} + c_i^\text{donor} \leq k_i^\text{donor} + 1$.
As an additional operationalisation of independent reinforcement, we count the number of different types of channels through which contagion can take place: family, professional, social, (former) colleagues, fellow alumni or fellow members, which we call \textit{source diversity}.
These measures are illustrated in Fig.~\ref{fig:network_example}.

To establish previous donation for candidates, we took into account all donations to all political candidates (senatorial, congressional, gubernatorial, presidential, et cetera) prior to the start of the election cycle.
For party donations, previous donations are more straightforward, and we simply check whether somebody donated  to the party prior to the start of the election cycle.

We use logistic regression, which is of the form
\begin{equation}
  \Pr(Y_i(t) = 1 \mid X_i(t - 1) = 0) = \frac{1}{1 + \exp (- \theta \mathbf{X})}
\end{equation}
where $\theta$ are parameters and $\mathbf{X}$ contains the covariates used in the specific model.
Since we pool the results for all quarters, we cluster errors on the donor for robust results.
Logistic regression is inherently similar to complex contagion and can be seen as a smooth approximation of a threshold model~\cite{Granovetter1978}.
Simple contagion would take another functional form.
Assuming each contagion is equally likely for each contact with some probability $p$, then the probability of infection with exposure to $k$ adopters is $1 - (1 - p)^k$ at each time step.
Fitting this model gave poor results, corroborating that the contagion of donations is not well modelled by simple contagion.

\section{Detailed results}
\label{sec:detailed_results}

\begin{figure*}[ht]
    \centering
    \includegraphics{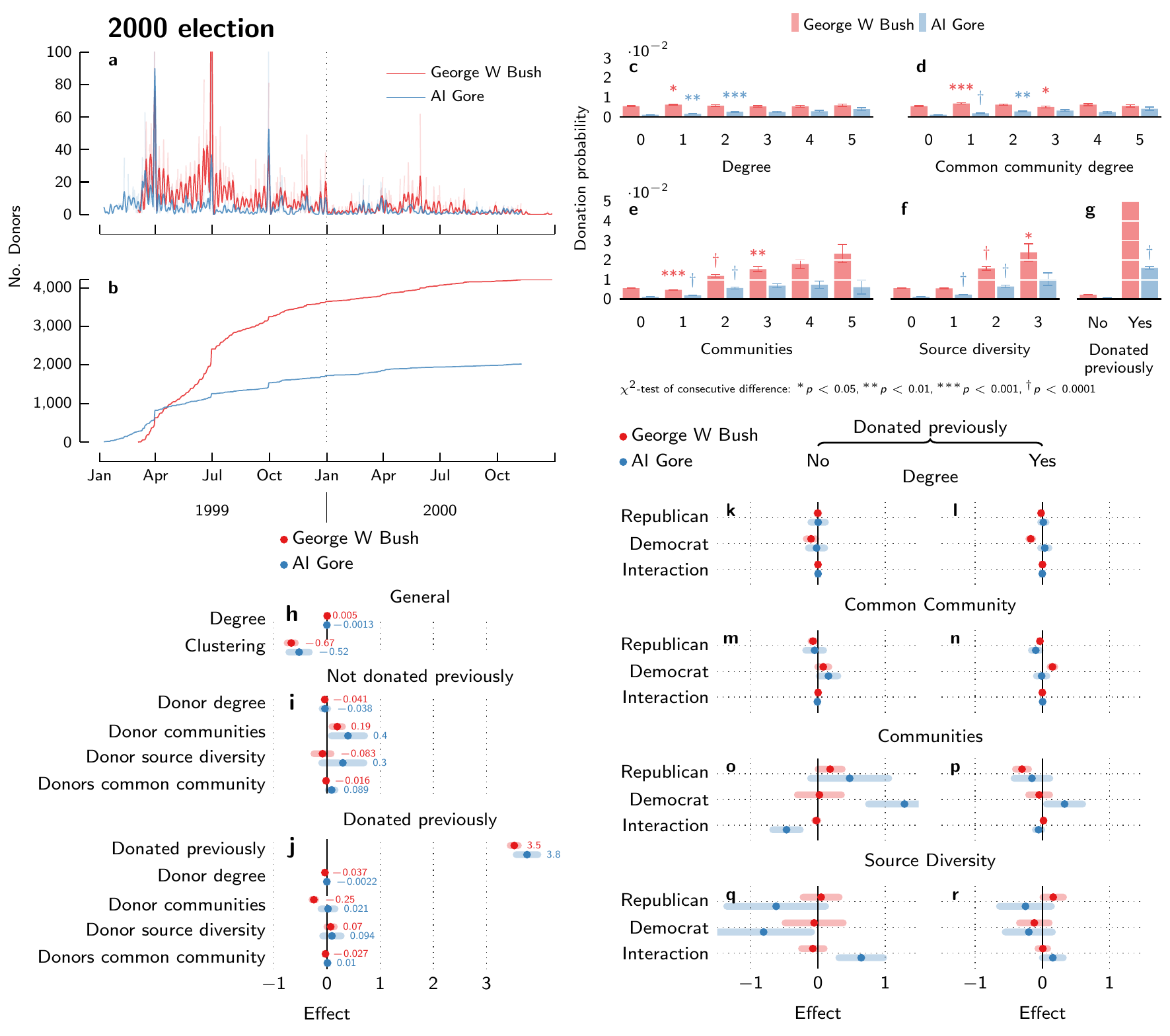}
    \caption{\textbf{Results for presidential candidates, election 2000.}
             Daily donors (a)---raw data is transparent, smoothed data is solid---and cumulative donors (b), probability effect of (c) donor degree, (d) common community donor degree, (e) donor communities, (f) source diversity and (g) previous donation.
             Logistic regression results (h) general effects, (i) network effects for new donors, and (j) network effects for old donors.
             Logistic regression result for cross-cutting effects, distinguished by old/new donors for donor degree (k)--(l), common community donor degree (m)--(n), donor communities (o)--(p), and source diversity (q)--(r).
             Error bars show 95\% confidence intervals for the coefficients.
`            }
    \label{SI_fig:presidential_2000}
\end{figure*}

\begin{figure*}[ht]
    \centering
    \includegraphics{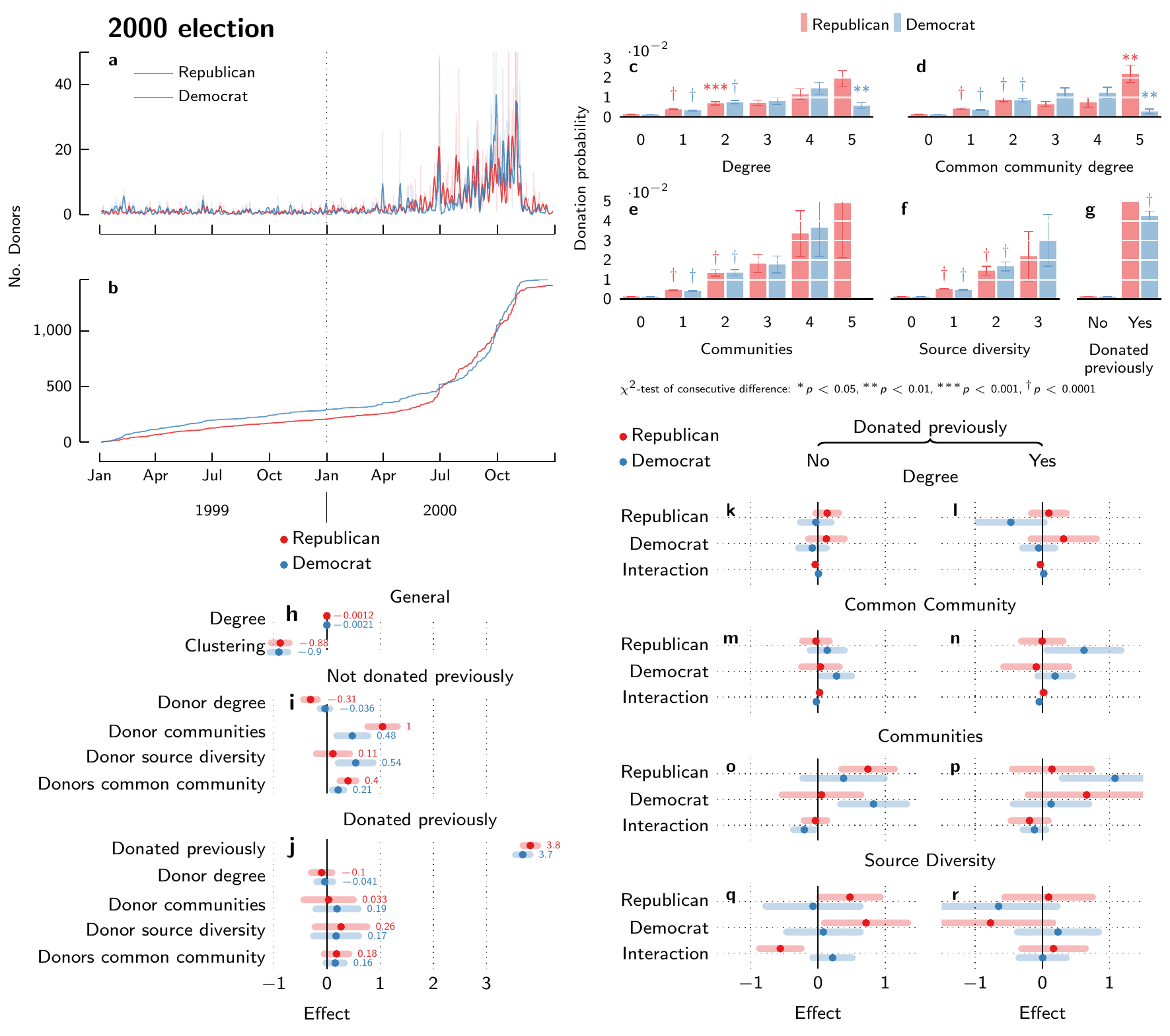}
    \caption{\textbf{Results for parties, election 2000.}
             Daily donors (a)---raw data is transparent, smoothed data is solid--- and cumulative donors (b), probability effect of (c) donor degree, (d) common community donor degree, (e) donor communities, (f) source diversity and (g) previous donation.
             Logistic regression results (h) general effects, (i) network effects for new donors, and (j) network effects for old donors.
             Logistic regression result for cross-cutting effects, distinguished by old/new donors for donor degree (k)--(l), common community donor degree (m)--(n), donor communities (o)--(p), and source diversity (q)--(r).
             Error bars show 95\% confidence intervals for the coefficients.}
    \label{SI_fig:party_2000}
\end{figure*}

\begin{figure*}[ht]
    \centering
    \includegraphics{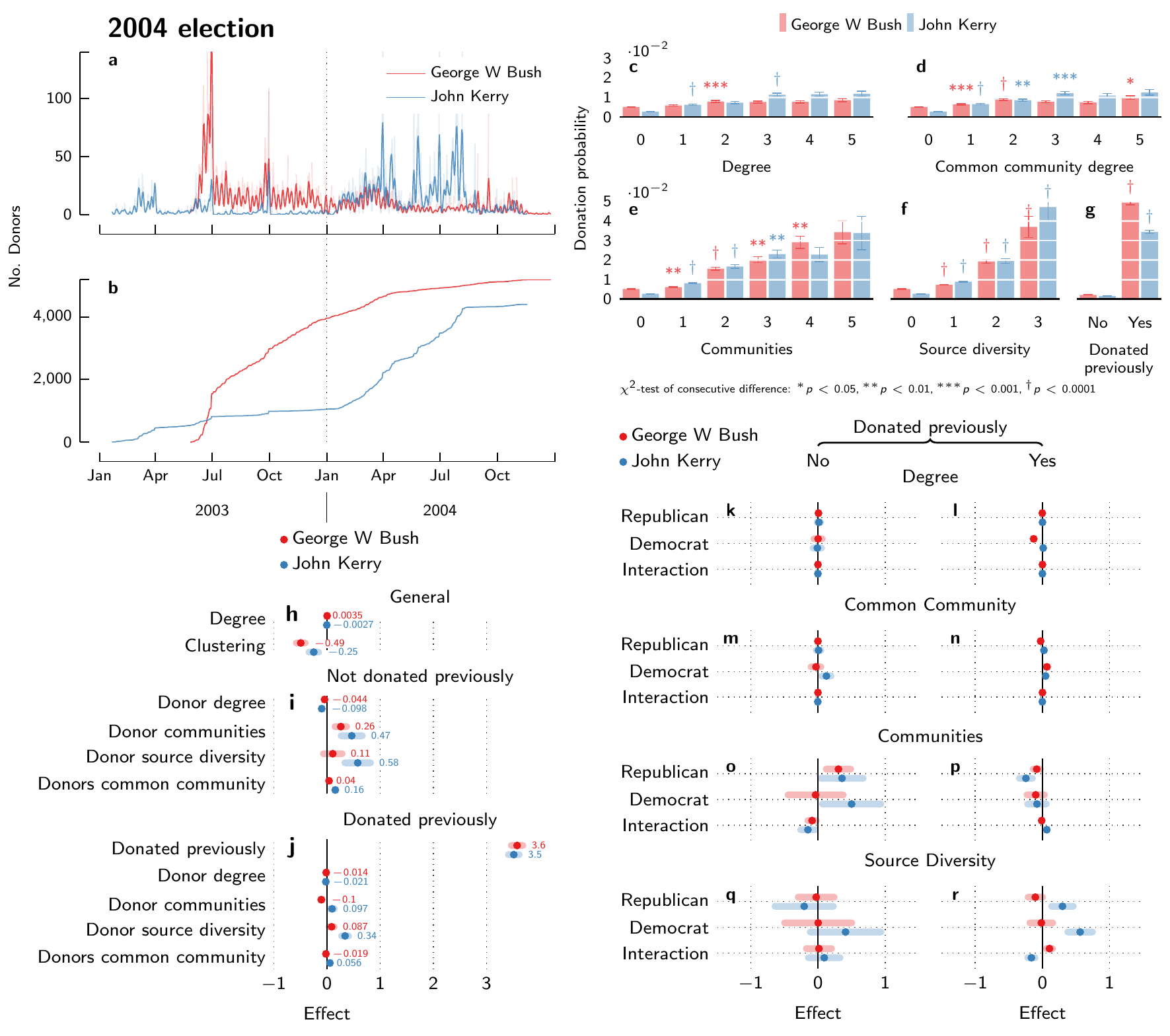}
    \caption{\textbf{Results for presidential candidates, election 2004.}
             Daily donors (a)---raw data is transparent, smoothed data is solid--- and cumulative donors (b), probability effect of (c) donor degree, (d) common community donor degree, (e) donor communities, (f) source diversity and (g) previous donation.
             Logistic regression results (h) general effects, (i) network effects for new donors, and (j) network effects for old donors.
             Logistic regression result for cross-cutting effects, distinguished by old/new donors for donor degree (k)--(l), common community donor degree (m)--(n), donor communities (o)--(p), and source diversity (q)--(r).
             Error bars show 95\% confidence intervals for the coefficients.}
    \label{SI_fig:presidential_2004}
\end{figure*}

\begin{figure*}[ht]
    \centering
    \includegraphics{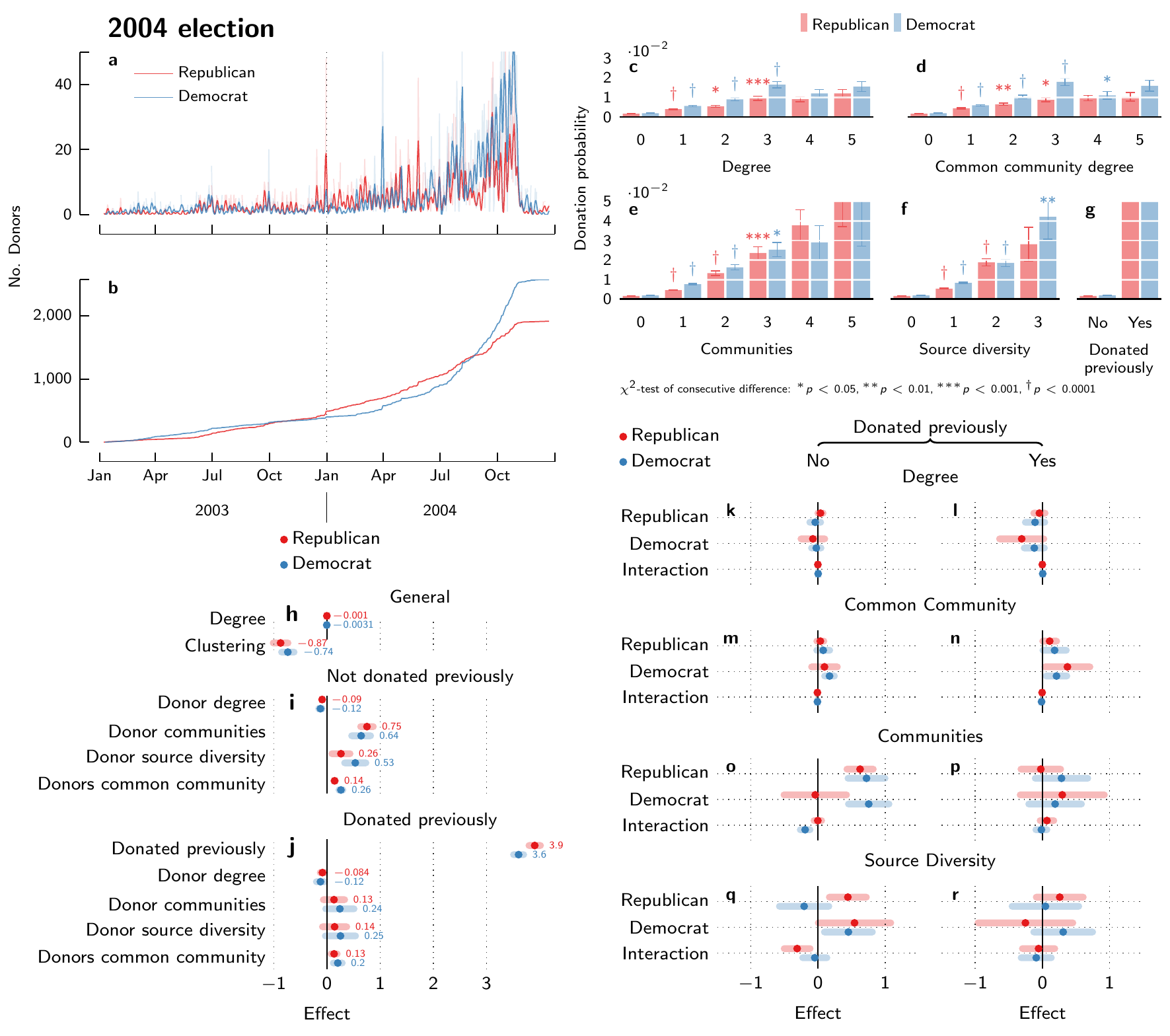}
    \caption{\textbf{Results for parties, election 2004.}
             Daily donors (a)---raw data is transparent, smoothed data is solid--- and cumulative donors (b), probability effect of (c) donor degree, (d) common community donor degree, (e) donor communities, (f) source diversity and (g) previous donation.
             Logistic regression results (h) general effects, (i) network effects for new donors, and (j) network effects for old donors.
             Logistic regression result for cross-cutting effects, distinguished by old/new donors for donor degree (k)--(l), common community donor degree (m)--(n), donor communities (o)--(p), and source diversity (q)--(r).
             Error bars show 95\% confidence intervals for the coefficients.}
    \label{SI_fig:party_2004}
\end{figure*}

\begin{figure*}[ht]
    \centering
    \includegraphics{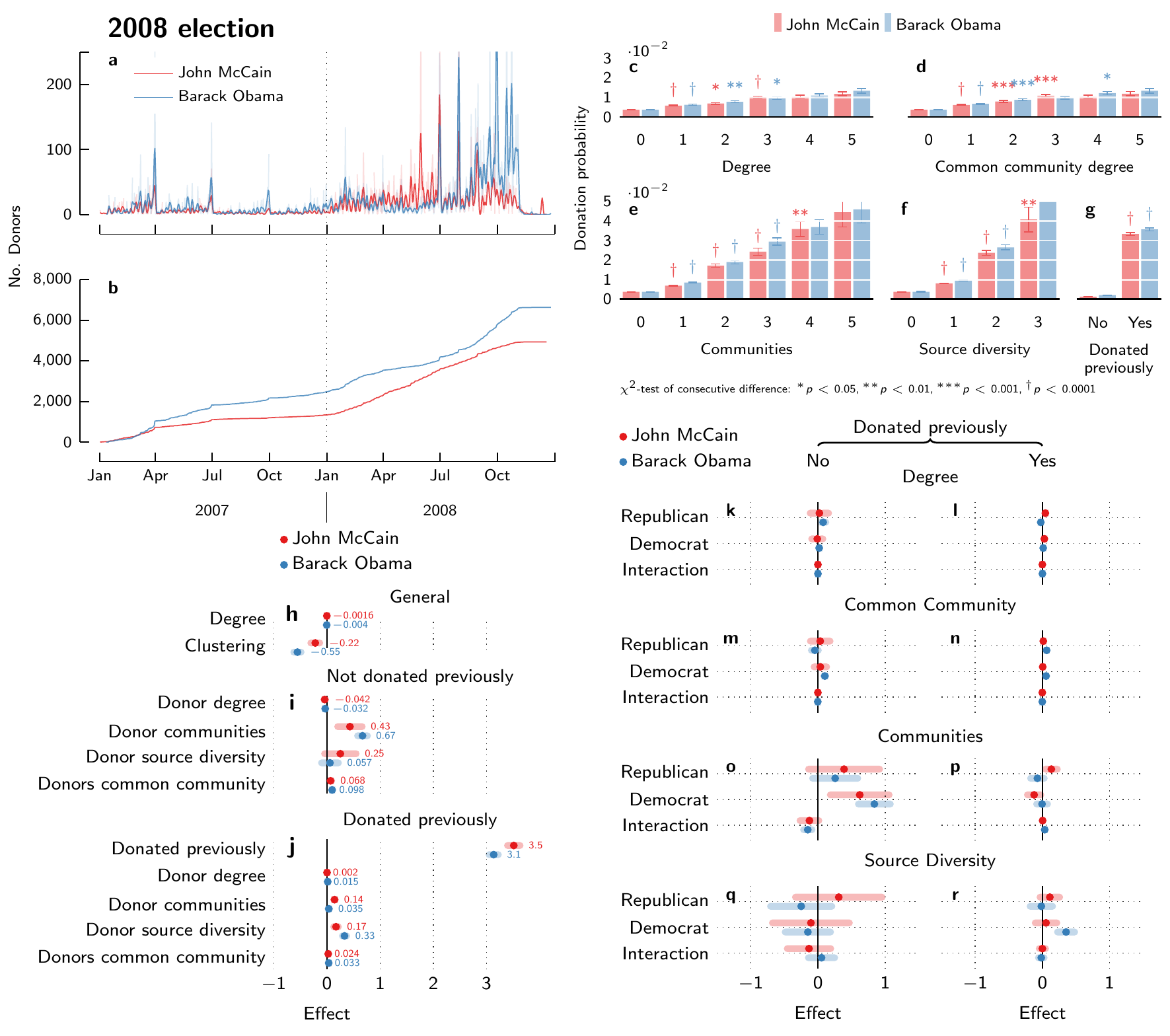}
    \caption{\textbf{Results for presidential candidates, election 2008.}
             Daily donors (a)---raw data is transparent, smoothed data is solid--- and cumulative donors (b), probability effect of (c) donor degree, (d) common community donor degree, (e) donor communities, (f) source diversity and (g) previous donation.
             Logistic regression results (h) general effects, (i) network effects for new donors, and (j) network effects for old donors.
             Logistic regression result for cross-cutting effects, distinguished by old/new donors for donor degree (k)--(l), common community donor degree (m)--(n), donor communities (o)--(p), and source diversity (q)--(r).
             Error bars show 95\% confidence intervals for the coefficients.}
    \label{SI_fig:presidential_2008}
\end{figure*}

\begin{figure*}[ht]
    \centering
    \includegraphics{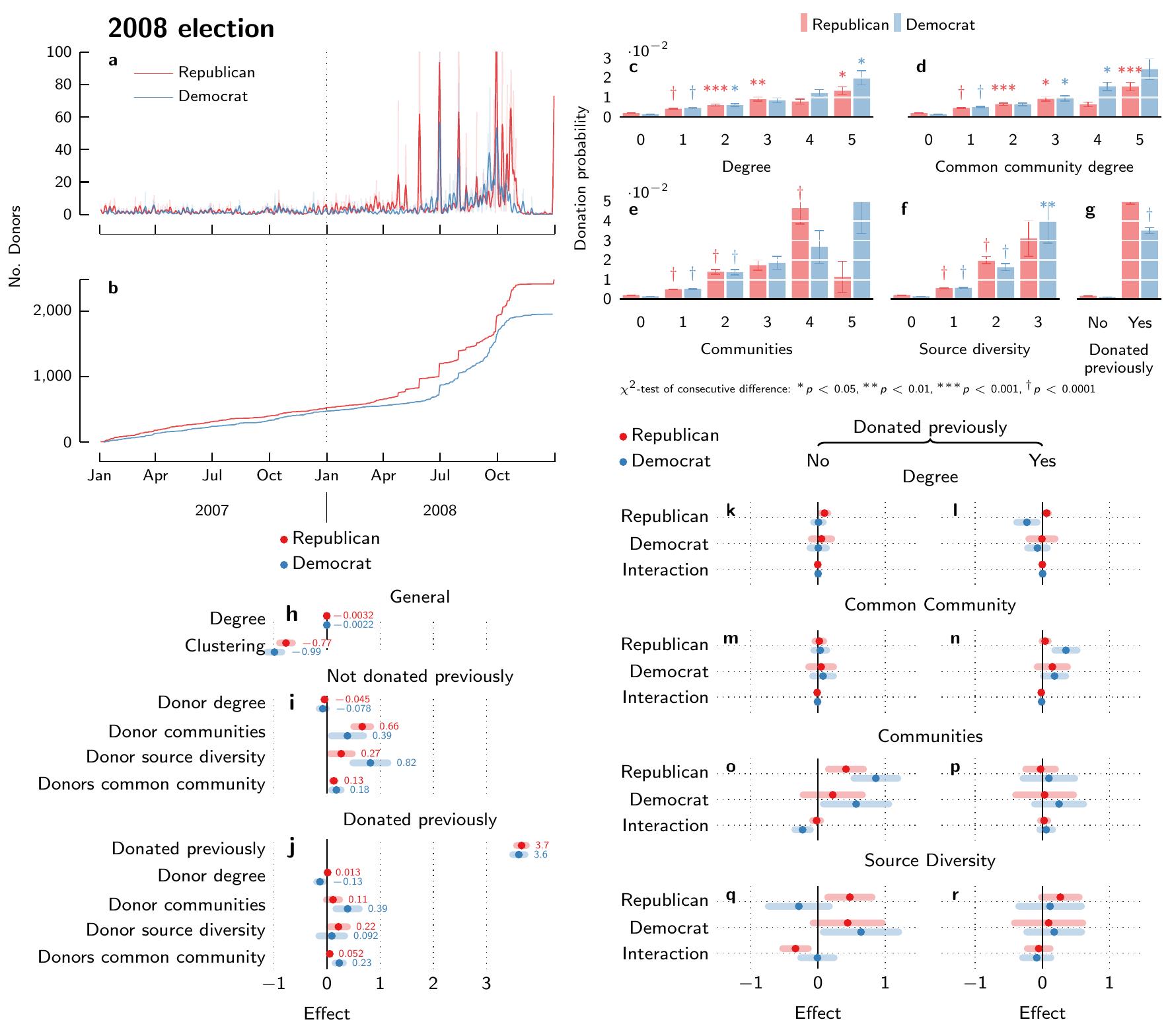}
    \caption{\textbf{Results for parties, election 2008.}
             Daily donors (a)---raw data is transparent, smoothed data is solid--- and cumulative donors (b), probability effect of (c) donor degree, (d) common community donor degree, (e) donor communities, (f) source diversity and (g) previous donation.
             Logistic regression results (h) general effects, (i) network effects for new donors, and (j) network effects for old donors.
             Logistic regression result for cross-cutting effects, distinguished by old/new donors for donor degree (k)--(l), common community donor degree (m)--(n), donor communities (o)--(p), and source diversity (q)--(r).
             Error bars show 95\% confidence intervals for the coefficients.}
    \label{SI_fig:party_2008}
\end{figure*}

\begin{figure*}[ht]
    \centering
    \includegraphics{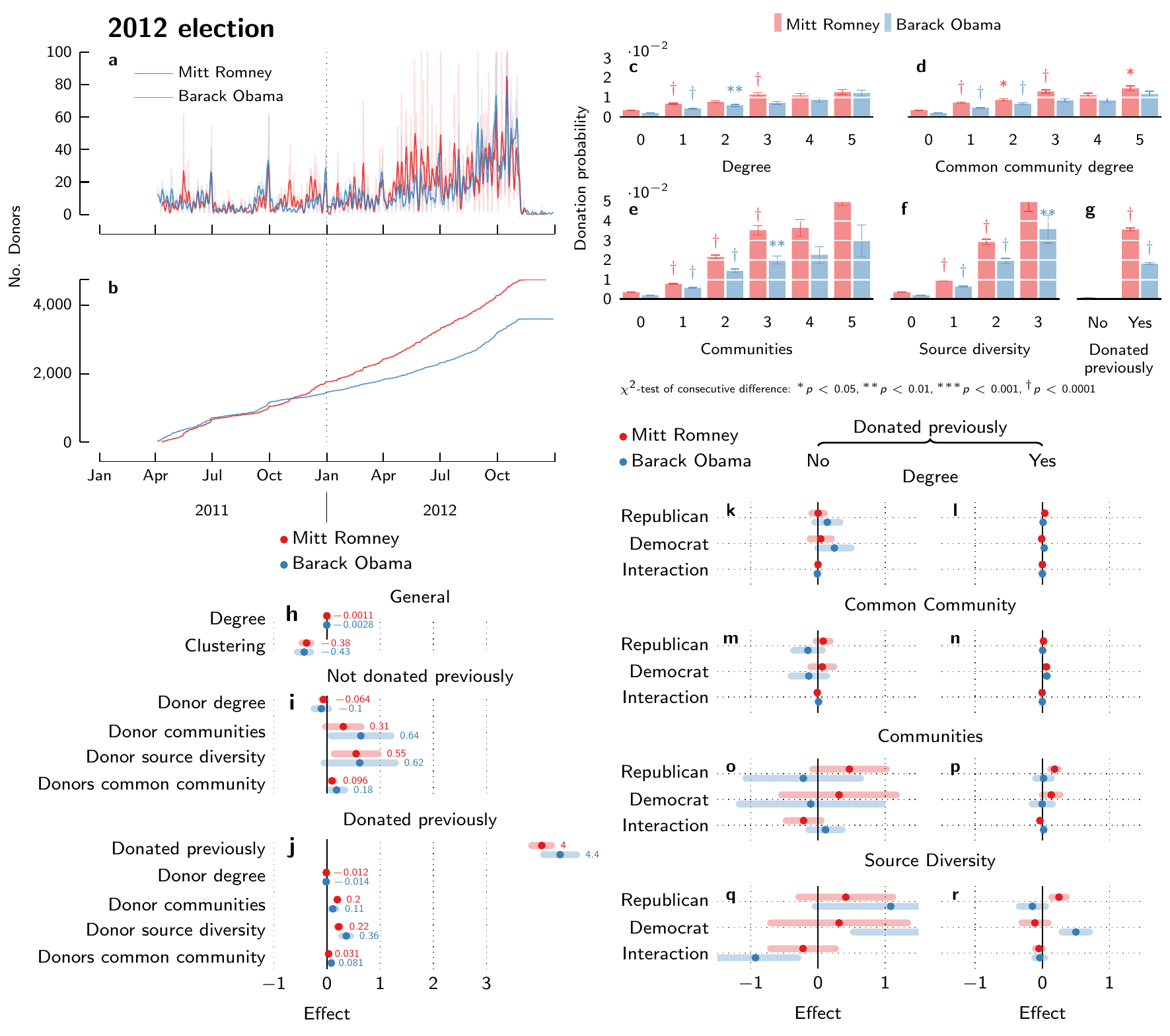}
    \caption{\textbf{Results for presidential candidates, election 2012.}
             Daily donors (a)---raw data is transparent, smoothed data is solid--- and cumulative donors (b), probability effect of (c) donor degree, (d) common community donor degree, (e) donor communities, (f) source diversity and (g) previous donation.
             Logistic regression results (h) general effects, (i) network effects for new donors, and (j) network effects for old donors.
             Logistic regression result for cross-cutting effects, distinguished by old/new donors for donor degree (k)--(l), common community donor degree (m)--(n), donor communities (o)--(p), and source diversity (q)--(r).
             Error bars show 95\% confidence intervals for the coefficients.}
    \label{SI_fig:presidential_2012}
\end{figure*}

\begin{figure*}[ht]
    \centering
    \includegraphics{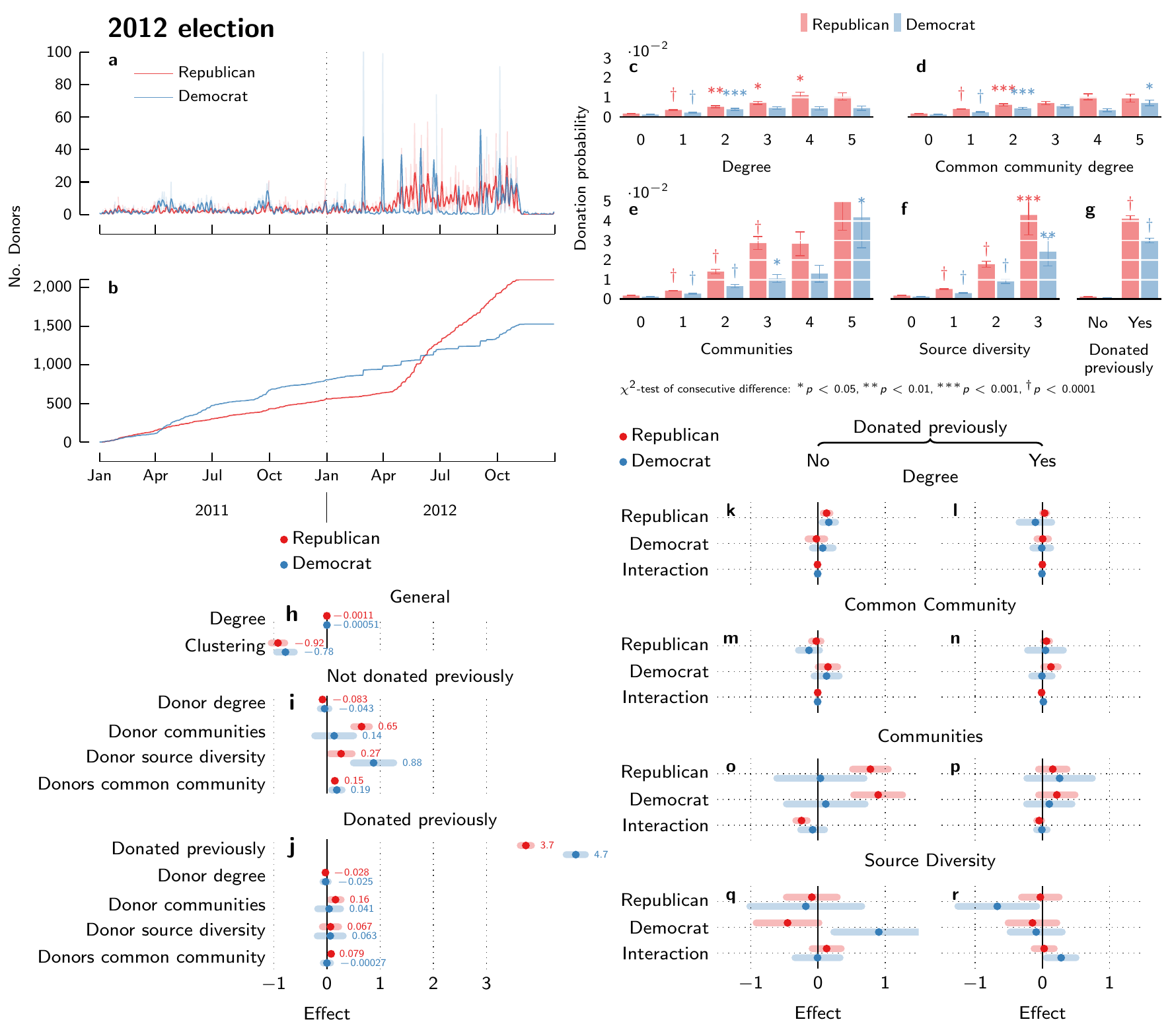}
    \caption{\textbf{Results for parties, election 2012}
             Daily donors (a)---raw data is transparent, smoothed data is solid--- and cumulative donors (b), probability effect of (c) donor degree, (d) common community donor degree, (e) donor communities, (f) source diversity and (g) previous donation.
             Logistic regression results (h) general effects, (i) network effects for new donors, and (j) network effects for old donors.
             Logistic regression result for cross-cutting effects, distinguished by old/new donors for donor degree (k)--(l), common community donor degree (m)--(n), donor communities (o)--(p), and source diversity (q)--(r).
             Error bars show 95\% confidence intervals for the coefficients.}
    \label{SI_fig:party_2012}
\end{figure*}

\begin{table*}
    \centering
\begin{tabular}{lllll}
\multicolumn{5}{r}{Dependent variable: donation to democratic candidate} \\
\toprule
                                                  Variable &                                      2000 &                                                2004 &                                                2008 &                                                2012 \\
\midrule
                                                 Intercept &         -7.5\textsuperscript{\textdagger} &                   -6.9\textsuperscript{\textdagger} &                   -6.2\textsuperscript{\textdagger} &                   -8.4\textsuperscript{\textdagger} \\
                                        Donated previously &          3.8\textsuperscript{\textdagger} &                    3.5\textsuperscript{\textdagger} &                    3.1\textsuperscript{\textdagger} &                    4.4\textsuperscript{\textdagger} \\
                                                    Degree &  $-1.3 \times 10^{-3}$\textsuperscript{*} &  $-2.7 \times 10^{-3}$\textsuperscript{\textdagger} &  $-4.0 \times 10^{-3}$\textsuperscript{\textdagger} &  $-2.8 \times 10^{-3}$\textsuperscript{\textdagger} \\
                                                Clustering &                -0.52\textsuperscript{***} &                           -0.25\textsuperscript{**} &                  -0.55\textsuperscript{\textdagger} &                  -0.43\textsuperscript{\textdagger} \\[1em]

                                              Donor degree &                                    -0.038 &                           -0.098\textsuperscript{*} &                           -0.032\textsuperscript{*} &                                               -0.10 \\
                  Donor degree $\times$ Donated previously &                                     0.036 &                                               0.077 &                           0.047\textsuperscript{**} &                                               0.091 \\
\end{tabular}

    \caption{Logistic regression results for 2000--2012 for donation to the democratic candidate as dependent variable.}
    \label{SI_tab:regression_candidates_D}
\end{table*}

\begin{table*}
    \centering
\begin{tabular}{lllll}
\multicolumn{5}{r}{Dependent variable: donation to republican candidate} \\
\toprule
                                                  Variable &                                               2000 &                                               2004 &                                                2008 &                                        2012 \\
\midrule
                                                 Intercept &                  -5.8\textsuperscript{\textdagger} &                  -6.2\textsuperscript{\textdagger} &                   -7.0\textsuperscript{\textdagger} &           -7.4\textsuperscript{\textdagger} \\
                                        Donated previously &                   3.5\textsuperscript{\textdagger} &                   3.6\textsuperscript{\textdagger} &                    3.5\textsuperscript{\textdagger} &            4.0\textsuperscript{\textdagger} \\
                                                    Degree &  $5.0 \times 10^{-3}$\textsuperscript{\textdagger} &  $3.5 \times 10^{-3}$\textsuperscript{\textdagger} &  $-1.6 \times 10^{-3}$\textsuperscript{\textdagger} &  $-1.1 \times 10^{-3}$\textsuperscript{***} \\
                                                Clustering &                 -0.67\textsuperscript{\textdagger} &                 -0.49\textsuperscript{\textdagger} &                           -0.22\textsuperscript{**} &          -0.38\textsuperscript{\textdagger} \\[1em]

                                              Donor degree &                          -0.041\textsuperscript{*} &                         -0.044\textsuperscript{**} &                                              -0.042 &                                      -0.064 \\
                  Donor degree $\times$ Donated previously &                               $4.0 \times 10^{-3}$ &                                              0.030 &                                               0.044 &                                       0.052 \\
\end{tabular}

    \caption{Logistic regression results for 2000--2012 for donation to the republican candidate as dependent variable.}
    \label{SI_tab:regression_candidates_R}
\end{table*}

\begin{table*}
    \centering
\begin{tabular}{lllll}
\multicolumn{5}{r}{Dependent variable: donation to democratic party} \\
\toprule
                                                  Variable &                                                2000 &                                                2004 &                                                2008 &                                2012 \\
\midrule
                                                 Intercept &                   -6.3\textsuperscript{\textdagger} &                   -6.0\textsuperscript{\textdagger} &                   -6.4\textsuperscript{\textdagger} &   -7.6\textsuperscript{\textdagger} \\
                                        Donated previously &                    3.7\textsuperscript{\textdagger} &                    3.6\textsuperscript{\textdagger} &                    3.6\textsuperscript{\textdagger} &    4.7\textsuperscript{\textdagger} \\
                                                    Degree &  $-2.1 \times 10^{-3}$\textsuperscript{\textdagger} &  $-3.1 \times 10^{-3}$\textsuperscript{\textdagger} &  $-2.2 \times 10^{-3}$\textsuperscript{\textdagger} &               $-5.1 \times 10^{-4}$ \\
                                                Clustering &                  -0.90\textsuperscript{\textdagger} &                  -0.74\textsuperscript{\textdagger} &                  -0.99\textsuperscript{\textdagger} &  -0.78\textsuperscript{\textdagger} \\[1em]
                                                
                                              Donor degree &                                              -0.036 &                                               -0.12 &                                              -0.078 &                              -0.043 \\
                  Donor degree $\times$ Donated previously &                               $-5.3 \times 10^{-3}$ &                                $1.1 \times 10^{-3}$ &                                              -0.053 &                               0.019 \\
\end{tabular}

    \caption{Logistic regression results for 2000--2012 for donation to the democratic party as dependent variable.}
    \label{SI_tab:regression_party_D}
\end{table*}

\begin{table*}
    \centering
\begin{tabular}{lllll}
\multicolumn{5}{r}{Dependent variable: donation to republican party} \\
\toprule
                                                  Variable &                                       2000 &                                       2004 &                                                2008 &                                       2012 \\
\midrule
                                                 Intercept &          -6.2\textsuperscript{\textdagger} &          -6.2\textsuperscript{\textdagger} &                   -6.1\textsuperscript{\textdagger} &          -6.4\textsuperscript{\textdagger} \\
                                        Donated previously &           3.8\textsuperscript{\textdagger} &           3.9\textsuperscript{\textdagger} &                    3.7\textsuperscript{\textdagger} &           3.7\textsuperscript{\textdagger} \\
                                                    Degree &  $-1.2 \times 10^{-3}$\textsuperscript{**} &  $-1.0 \times 10^{-3}$\textsuperscript{**} &  $-3.2 \times 10^{-3}$\textsuperscript{\textdagger} &  $-1.1 \times 10^{-3}$\textsuperscript{**} \\
                                                Clustering &         -0.88\textsuperscript{\textdagger} &         -0.87\textsuperscript{\textdagger} &                  -0.77\textsuperscript{\textdagger} &         -0.92\textsuperscript{\textdagger} \\[1em]
                                                
                                              Donor degree &                 -0.31\textsuperscript{***} &                  -0.090\textsuperscript{*} &                                              -0.045 &                  -0.083\textsuperscript{*} \\
                  Donor degree $\times$ Donated previously &                                       0.21 &                       $6.0 \times 10^{-3}$ &                                               0.058 &                                      0.055 \\
\end{tabular}

    \caption{Lofiguregistic regression results for 2000--2012 for donation to the republican party as dependent variable.}
    \label{SI_tab:regression_party_R}
\end{table*}
\begin{table*}
    \centering
\begin{tabular}{lllllll}
\multicolumn{7}{r}{Dependent variable: donation to democratic candidate} \\
\toprule
                      \multicolumn{3}{l}{Variable}                                                                         &                                      2000 &                                                2004 &                                                2008 &                                                2012 \\
\midrule
                     \multicolumn{3}{l}{Intercept}                                                                         &         -7.6\textsuperscript{\textdagger} &                   -6.9\textsuperscript{\textdagger} &                   -6.2\textsuperscript{\textdagger} &                   -8.8\textsuperscript{\textdagger} \\
                        \multicolumn{3}{l}{Degree}                                                                         &  $1.8 \times 10^{-3}$\textsuperscript{**} &  $-2.8 \times 10^{-3}$\textsuperscript{\textdagger} &  $-4.9 \times 10^{-3}$\textsuperscript{\textdagger} &  $-2.6 \times 10^{-3}$\textsuperscript{\textdagger} \\
                    \multicolumn{3}{l}{Clustering}                                                                         &                 -0.46\textsuperscript{**} &                  -0.35\textsuperscript{\textdagger} &                  -0.64\textsuperscript{\textdagger} &                  -0.47\textsuperscript{\textdagger} \\
            \multicolumn{3}{l}{Donated previously (D)}                                                                     &          4.1\textsuperscript{\textdagger} &                    3.7\textsuperscript{\textdagger} &                    3.2\textsuperscript{\textdagger} &                    4.9\textsuperscript{\textdagger} \\[1em]
                  \multicolumn{3}{l}{Donor degree} \\
\end{tabular}

    \caption{Logistic regression results for 2000--2012 for donation to the democratic candidate as dependent variable, including cross-exposure effects (i.e. effect of exposure to democratic donors on republican donation).}
    \label{SI_tab:regression_cross_candidates_D}
\end{table*}

\begin{table*}
    \centering
\begin{tabular}{lllllll}
\multicolumn{7}{r}{Dependent variable: donation to republican candidate} \\
\toprule
                      \multicolumn{3}{l}{Variable}                                                                         &                                               2000 &                                               2004 &                                                2008 &                                        2012 \\
\midrule
                     \multicolumn{3}{l}{Intercept}                                                                         &                  -5.8\textsuperscript{\textdagger} &                  -6.1\textsuperscript{\textdagger} &                   -7.2\textsuperscript{\textdagger} &           -7.7\textsuperscript{\textdagger} \\
                        \multicolumn{3}{l}{Degree}                                                                         &  $7.0 \times 10^{-3}$\textsuperscript{\textdagger} &  $4.7 \times 10^{-3}$\textsuperscript{\textdagger} &  $-2.5 \times 10^{-3}$\textsuperscript{\textdagger} &  $-1.3 \times 10^{-3}$\textsuperscript{***} \\
                    \multicolumn{3}{l}{Clustering}                                                                         &                 -0.74\textsuperscript{\textdagger} &                 -0.50\textsuperscript{\textdagger} &                           -0.25\textsuperscript{**} &          -0.44\textsuperscript{\textdagger} \\
            \multicolumn{3}{l}{Donated previously (R)}                                                                     &                   3.6\textsuperscript{\textdagger} &                   3.8\textsuperscript{\textdagger} &                    3.8\textsuperscript{\textdagger} &            4.3\textsuperscript{\textdagger} \\[1em]
                  \multicolumn{3}{l}{Donor degree} \\
\end{tabular}

    \caption{Logistic regression results for 2000--2012 for donation to the republican candidate as dependent variable, including cross-exposure effects (i.e. effect of exposure to democratic donors on republican donation).}
    \label{SI_tab:regression_cross_candidates_R}
\end{table*}

\begin{table*}
    \centering
\begin{tabular}{lllllll}
\multicolumn{7}{r}{Dependent variable: donation to democratic party} \\
\toprule
                      \multicolumn{3}{l}{Variable}                                                                         &                                                2000 &                                                2004 &                                                2008 &                                2012 \\
\midrule
                     \multicolumn{3}{l}{Intercept}                                                                         &                   -6.3\textsuperscript{\textdagger} &                   -6.2\textsuperscript{\textdagger} &                   -6.6\textsuperscript{\textdagger} &   -7.7\textsuperscript{\textdagger} \\
                        \multicolumn{3}{l}{Degree}                                                                         &  $-2.9 \times 10^{-3}$\textsuperscript{\textdagger} &  $-4.2 \times 10^{-3}$\textsuperscript{\textdagger} &  $-3.5 \times 10^{-3}$\textsuperscript{\textdagger} &                $8.1 \times 10^{-5}$ \\
                    \multicolumn{3}{l}{Clustering}                                                                         &                  -0.89\textsuperscript{\textdagger} &                  -0.71\textsuperscript{\textdagger} &                  -0.98\textsuperscript{\textdagger} &  -0.73\textsuperscript{\textdagger} \\
            \multicolumn{3}{l}{Donated previously (D)}                                                                     &                    3.7\textsuperscript{\textdagger} &                    3.7\textsuperscript{\textdagger} &                    3.7\textsuperscript{\textdagger} &    4.8\textsuperscript{\textdagger} \\[1em]
                  \multicolumn{3}{l}{Donor degree} \\
\end{tabular}

    \caption{Logistic regression results for 2000--2012 for donation to the democratic party as dependent variable, including cross-exposure effects (i.e. effect of exposure to republican donors on democratic donation).}
    \label{SI_tab:regression_cross_party_D}
\end{table*}

\begin{table*}
    \centering
\begin{tabular}{lllllll}
\multicolumn{7}{r}{Dependent variable: donation to republican party} \\
\toprule
                                          \multicolumn{3}{l}{Variable}                                                     &                                                2000 &                                       2004 &                                                2008 &                                                2012 \\
\midrule
                     \multicolumn{3}{l}{Intercept}                                                                         &                   -6.3\textsuperscript{\textdagger} &          -6.3\textsuperscript{\textdagger} &                   -6.2\textsuperscript{\textdagger} &                   -6.6\textsuperscript{\textdagger} \\
                        \multicolumn{3}{l}{Degree}                                                                         &  $-2.2 \times 10^{-3}$\textsuperscript{\textdagger} &  $-1.0 \times 10^{-3}$\textsuperscript{**} &  $-3.9 \times 10^{-3}$\textsuperscript{\textdagger} &  $-2.1 \times 10^{-3}$\textsuperscript{\textdagger} \\
                    \multicolumn{3}{l}{Clustering}                                                                         &                  -0.88\textsuperscript{\textdagger} &         -0.90\textsuperscript{\textdagger} &                  -0.78\textsuperscript{\textdagger} &                  -0.88\textsuperscript{\textdagger} \\
            \multicolumn{3}{l}{Donated previously (R)}                                                                     &                    4.0\textsuperscript{\textdagger} &           4.0\textsuperscript{\textdagger} &                    3.8\textsuperscript{\textdagger} &                    3.9\textsuperscript{\textdagger} \\[1em]
                  \multicolumn{3}{l}{Donor degree} \\
\end{tabular}

    \caption{Logistic regression results for 2000--2012 for donation to the republican party as dependent variable, including cross-exposure effects (i.e. effect of exposure to democratic donors on republican donation).}
    \label{SI_tab:regression_cross_party_R}
\end{table*}
\end{document}